\documentclass[12pt,a4paper]{article}
\usepackage{amsmath}
\usepackage{amssymb}
\usepackage{slashed}
\usepackage[pdftex]{graphicx}
\usepackage{listings}
\usepackage[colorlinks,citecolor=blue,urlcolor=blue,linkcolor=blue]{hyperref}

\newcommand{\GeV}{\mbox{$~{\rm GeV}$}}
\newcommand{\TeV}{\mbox{$~{\rm TeV}$}}
\newcommand{\meg}{\mu\to e\gamma}
\newcommand{\mte}{\mu\to 3 \, e}
\newcommand{\mec}{\mu\mbox{-}e}
\newcommand{\mdm}{M_{N_1}}
\newcommand{\mch}{m_{\eta^+}}

\title{\bf Probing the scotogenic model with lepton flavor violating processes}
\author{Avelino Vicente\\
{\it\small IFPA, Dep. AGO, Universit\'e Li\`ege,}\\
{\it\small Bat B5, Sart-Tilman B-4000, Li\`ege 1, Belgium}\\
 \and Carlos E. Yaguna\\{\it \small Institut f\"ur Theoretische Physik, Universit\"at M\"unster,}\\
{\it \small Wilhelm-Klemm-Stra\ss e 9, D-48149 M\"unster, Germany}
}
\date{}

\begin{document}

\maketitle
\vspace*{-10cm}
\begin{flushright}
\texttt{\footnotesize MS-TP-14-37}
\end{flushright}
\vspace*{8.5cm}
\thispagestyle{empty}
\begin{abstract}
We study the impact that future lepton flavor violating experiments will have on the viable parameter space of the scotogenic model.  Within this model, the dark matter particle is assumed to be the lightest singlet fermion and two cases are considered depending on how its relic density is obtained: via self-annihilations or via coannihilations with the scalars. For each case, a scan over the parameter space of the model is used to obtain a large sample of viable points, which we subsequently analyze. We find that future lepton flavor violating experiments, in particular those searching for $\mte$ and $\mec$ conversion in nuclei, will probe the parameter space of the  scotogenic model in a significant way. They may exclude a large fraction of the models where the dark matter density is determined by coannihilations, and could rule out all the models where it is determined by annihilations.    
\end{abstract}

\section{Motivation}
\label{sec:motivation}
The experimental bounds on lepton flavor violating processes will be greatly improved in the near future \cite{Mihara:2013zna}. For  lepton flavor violating $\tau$ decays, such as $\tau\to\ell\gamma$ and $\tau\to 3 \, \ell$, the expected future sensitivity is about one order of magnitude below  their present limits, which already exclude branching ratios larger than about $10^{-8}$.   In the $\mec$ sector, current limits are more stringent and the expected improvements are more significant. For $\mte$ a sensitivity four orders of magnitude below the present bound is foreseen, while the limit on $\mec$ conversion in nuclei  could be increased by up to  six orders of magnitude. Even for $\meg$, which currently provides the strongest bound, a one order of magnitude improvement is expected in the near future.  Given that the present limits on some of these processes are already very impressive and can restrict the parameter space of new physics models in an important way, one can only wonder about the impact that these future experimental improvements might have on such models. Could they exclude some scenarios? How will they affect their viable parameter space? In this paper, we address precisely these issues within a specific and well-motivated extension of the Standard Model: the scotogenic model.

The scotogenic model \cite{Ma:2006km} is probably the simplest TeV scale model that can simultaneously account for neutrino masses and dark matter. It contains  three additional singlet fermions, $N_i$ ($i=1,2,3$), and another scalar doublet, $\eta$, all assumed to be odd under a $\mathbb{Z}_2$ symmetry.  Neutrino masses are generated via 1-loop diagrams mediated by the odd particles, whereas the dark matter candidate is either the lightest singlet fermion or the neutral component of the scalar doublet. In this model, lepton flavor violating processes, such as $\meg$, $\tau\to\mu\gamma$, and $\mte$, take place at 1-loop, via diagrams analogous to those responsible for neutrino masses. In fact, it is well-known that the current experimental bounds on these processes, particularly $\meg$, already restrict its viable parameter space \cite{Toma:2013zsa}. The scotogenic model thus provides the perfect scenario to assess the impact of future lepton flavor violation experiments.

Lepton flavor violation in the scotogenic model has already been studied in some detail. Early works, including \cite{Kubo:2006yx,Sierra:2008wj,Suematsu:2009ww,Adulpravitchai:2009gi}, focused almost exclusively on $\meg$, due to its stringent experimental limit.  Other processes, such as $\mte$ or $\mec$ conversion in nuclei, were rarely considered, and  when they were so, only  the photonic dipole contribution was taken into account. This situation was recently remedied in \cite{Toma:2013zsa}, where complete analytical expressions for the most important lepton flavor violating (LFV) processes were obtained. But a thorough analysis of lepton flavor violation in this model including the constraints from the dark matter density and the expected improvements in LFV experiments was yet to be done.

To that end, we first randomly scan the entire parameter space of this model and find a large sample of points consistent with all current bounds, particularly neutrino masses, $\meg$, and dark matter. We focus on the scenario where the dark matter particle is the singlet fermion, $N_1$, and distinguish two relevant cases depending on the processes that determine its relic density in the early Universe:  $N_1$-$N_1$ annihilations, or $N_1$-$\eta$ coannihilations. For each case, the sample of consistent models defines the viable parameter space, which we analyze in detail. Then, we study the predictions for different lepton flavor violating processes within these viable regions, and examine to which extent future experiments will be able to probe them.  We will show that future LFV experiments have the potential to rule-out the entire parameter space consistent with a relic density determined by $N_1$-$N_1$ annihilations, and to exclude a significant part of that determined by  $N_1$-$\eta$ coannihilations.

In the next section, we describe the scotogenic model and introduce our notation.  Section \ref{sec:lfv} deals with lepton flavor violating processes. After reporting the experimental situation, we review the expressions for the rates of different LFV processes in the scotogenic model . Our main results are presented in  section \ref{sec:res}, first for the case where the  dark matter density is obtained without coannihilations and then with coannihilations. In that section, we thoroughly  discuss the viable parameter space and the impact of future LFV experiments. Finally, we summarize our findings in section \ref{sec:con}. 
\section{The model}
\label{sec:model}

The \emph{scotogenic model}~\cite{Ma:2006km} extends the SM particle content with three singlet fermions, $N_i$ ($i=1$-$3$), and one $SU(2)_L$ doublet, $\eta$. In addition, a $\mathbb{Z}_2$ parity is imposed, under which the new particles are odd and the SM ones are even. This symmetry not only prevents flavor changing neutral currents but it also renders stable the lightest odd particle in the spectrum, which  becomes a dark matter candidate. In this model, two particles can play the role of dark matter: the neutral scalar (an inert Higgs) or the lightest singlet fermion. Both have been shown to give rise to  interesting dark matter scenarios \cite{Kubo:2006yx,Sierra:2008wj,Suematsu:2009ww,Gelmini:2009xd,Schmidt:2012yg,Kashiwase:2012xd,Kashiwase:2013uy,Klasen:2013jpa}. Additionally, this model may also   generate new signals at colliders \cite{Sierra:2008wj,Aoki:2010tf,Gustafsson:2012aj,Ho:2013hia,Arhrib:2013ela}, explain the baryon asymmetry of the Universe \cite{Hambye:2009pw,Racker:2013lua}, induce observable rates for lepton flavor violating processes \cite{Toma:2013zsa}, and  account for the observed pattern of neutrino masses and mixing angles \cite{Ma:2006km,Ma:2012if}.

The new Lagrangian terms involving the right-handed neutrinos can be written as
\begin{equation}
\mathcal{L}_N=\overline{N_i}\partial\!\!\!/N_i
-\frac{M_{N_i}}{2}\overline{N_i^c}P_RN_i+
y_{i\alpha}\eta\overline{N_i}P_L\ell_\alpha+\mathrm{h.c.} \, ,
\end{equation}
where, without loss of generality, the right-handed neutrino mass matrix has been taken to be diagonal. The  matrix of Yukawa couplings, $y$, is an arbitrary $3 \times 3$ complex  matrix. The scalar potential of the model is given by 
\begin{eqnarray}
\mathcal{V}\!\!\!&=&\:
m_{\phi}^2\phi^\dag\phi+m_\eta^2\eta^\dag\eta+
\frac{\lambda_1}{2}\left(\phi^\dag\phi\right)^2+
\frac{\lambda_2}{2}\left(\eta^\dag\eta\right)^2+
\lambda_3\left(\phi^\dag\phi\right)\left(\eta^\dag\eta\right)\nonumber\\
&&\!\!\!+
\lambda_4\left(\phi^\dag\eta\right)\left(\eta^\dag\phi\right)+
\frac{\lambda_5}{2}\left[\left(\phi^\dag\eta\right)^2+
\left(\eta^\dag\phi\right)^2\right] \, .
\end{eqnarray}
For simplicity, we will assume that all parameters in the scalar potential are real. This, however, can always be accomplished by making use of the rephasing invariance of the model.  In the scotogenic model, the $\mathbb{Z}_2$ parity is assumed to be preserved after electroweak symmetry breaking. This is guaranteed by choosing a set of parameters that leads to a vacuum with $\langle \eta \rangle = 0$ and forbids the $\phi-\eta$ mixing.

After electroweak symmetry breaking, the masses of the charged component $\eta^+$ and neutral component $\eta^0=(\eta_R+i\eta_I)/\sqrt{2}$ are split to
\begin{eqnarray}
m_{\eta^+}^2&=&m_\eta^2+\lambda_3\langle\phi^0\rangle^2 \, , \label{eq:mass1}\\
m_R^2&=&m_{\eta}^2+\left(\lambda_3+\lambda_4+\lambda_5\right)\langle\phi^0\rangle^2 \, ,\label{eq:mass2}\\
m_I^2&=&m_{\eta}^2+\left(\lambda_3+\lambda_4-\lambda_5\right)\langle\phi^0\rangle^2 \, . \label{eq:mass3}
\end{eqnarray}
The mass difference between $\eta_R$ and $\eta_I$ (the CP-even and CP-odd components of $\eta^0$, respectively) is $m_R^2-m_I^2=2\lambda_5\langle\phi^0\rangle^2$.

\begin{figure}[t]
\centering
\includegraphics[scale=0.5]{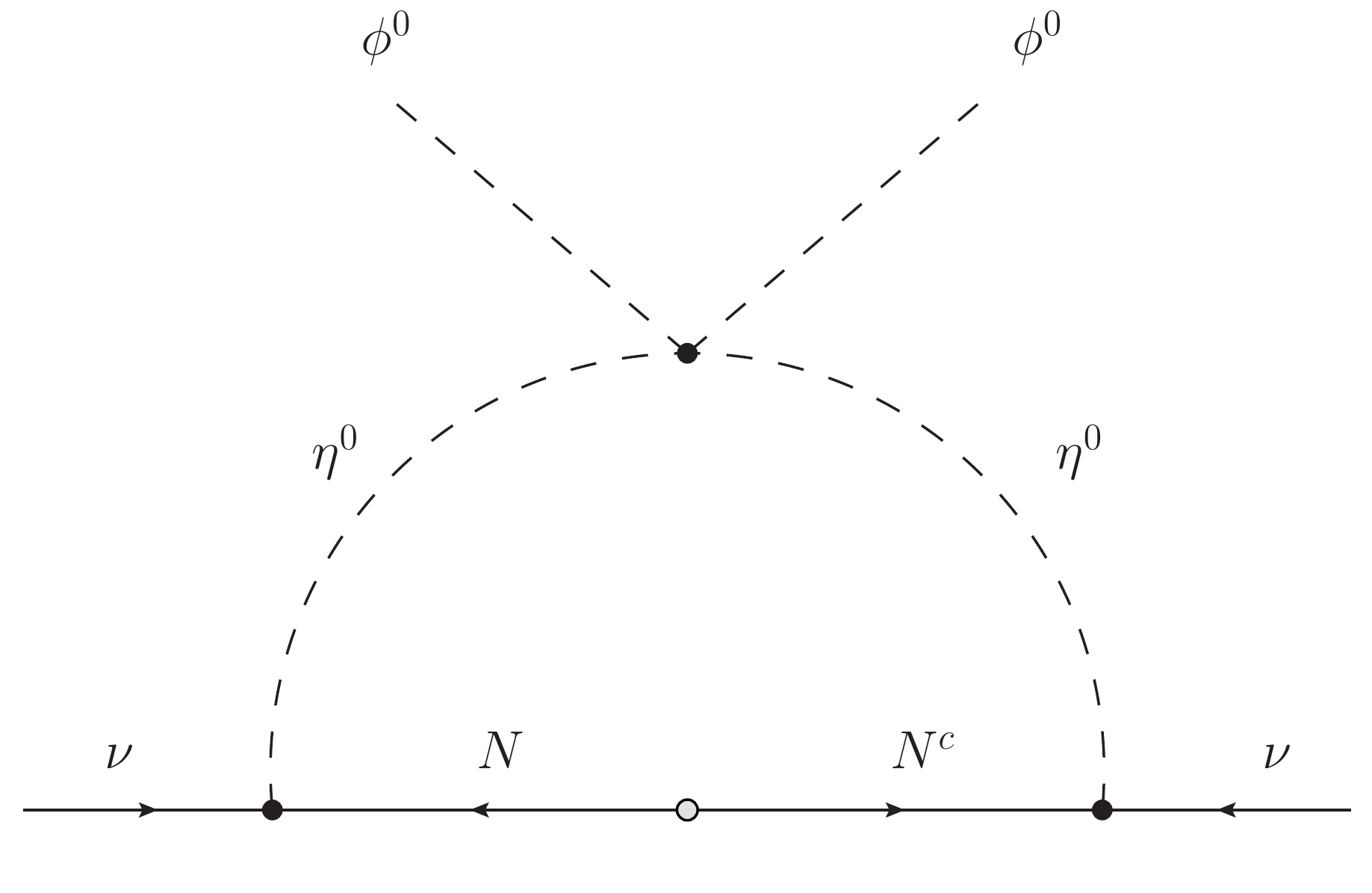}
\caption{1-loop neutrino masses in the scotogenic model. \label{fig:numass}}
\end{figure}
Inspecting the new terms in $\mathcal{L}_N$ and $\mathcal{V}$ one finds that the presence of $\lambda_5 \neq 0$ breaks lepton number in two units. Although the usual tree-level contribution to neutrino masses is forbidden by the $\mathbb{Z}_2$ symmetry, these are induced at the 1-loop level as shown in figure \ref{fig:numass}. This loop is calculable and leads to the neutrino mass matrix
\begin{eqnarray}
\left(m_{\nu}\right)_{\alpha\beta}&=&
\sum_{i=1}^3\frac{y_{i\alpha}y_{i\beta}}{(4\pi)^2} M_{N_i}
\left[\frac{m_R^2}{m_R^2-M_{N_i}^2}\log\left(\frac{m_R^2}{M_{N_i}^2}\right)
-\frac{m_I^2}{m_I^2-M_{N_i}^2}\log\left(\frac{m_I^2}{M_{N_i}^2}\right)\right]\nonumber\\
&\equiv&\left(y^{T}\Lambda y\right)_{\alpha\beta},
\label{eq:nu-mass}
\end{eqnarray}
where the $\Lambda$ matrix is defined as $\Lambda = \text{diag}\left( \Lambda_1,\Lambda_2,\Lambda_3 \right)$, with
\begin{equation}
\Lambda_i=\frac{M_{N_i}}{(4\pi)^2}
\left[\frac{m_R^2}{m_R^2-M_{N_i}^2}\log\left(\frac{m_R^2}{M_{N_i}^2}\right)
-\frac{m_I^2}{m_I^2-M_{N_i}^2}\log\left(\frac{m_I^2}{M_{N_i}^2}\right)\right] \, . 
\end{equation}
Simplified expressions can be obtained when $m_R^2 \approx m_I^2 \equiv m_0^2$ ($\lambda_5\ll1$). In this case the mass matrix in equation \eqref{eq:nu-mass} can be written as
\begin{equation} \label{eq:nu-mass2}
\left(m_\nu\right)_{\alpha\beta}\approx
\sum_{i=1}^3\frac{2\lambda_5 y_{i\alpha}y_{i\beta}\langle\phi^0\rangle^2}
{(4\pi)^2 M_{N_i}}
\left[\frac{M_{N_i}^2}{m_{0}^2-M_{N_i}^2}
+\frac{M_{N_i}^4}{\left(m_{0}^2-M_{N_i}^2\right)^2}
\log\left(\frac{M_{N_i}^2}{m_0^2}\right)\right] \, .
\end{equation}
Compared to the standard seesaw formula, neutrino masses get an additional suppression by roughly the factor $\sim \lambda_5 / 16 \pi^2$. Choosing $\lambda_5 \ll 1$, one can get the correct size for neutrino masses, compatible with singlet fermions at the TeV scale (or below) and sizable Yukawa couplings.

The neutrino mass matrix in equation \eqref{eq:nu-mass2} is diagonalized as
\begin{equation}
U_{\mathrm{PMNS}}^{T} \, m_{\nu} \, U_{\mathrm{PMNS}}=\hat{m}_{\nu}\equiv
\left(
\begin{array}{ccc}
m_1 & 0 & 0\\
0 & m_2 & 0\\
0 & 0 & m_3
\end{array}
\right) \, ,
\end{equation}
where 
\begin{equation}
\label{eq:PMNS}
U_{\mathrm{PMNS}}=
\left(
\begin{array}{ccc}
 c_{12}c_{13} & s_{12}c_{13}  & s_{13}e^{i\delta}  \\
-s_{12}c_{23}-c_{12}s_{23}s_{13}e^{-i\delta}  & 
c_{12}c_{23}-s_{12}s_{23}s_{13}e^{-i\delta}  & s_{23}c_{13}  \\
s_{12}s_{23}-c_{12}c_{23}s_{13}e^{-i\delta}  & 
-c_{12}s_{23}-s_{12}c_{23}s_{13}e^{-i\delta}  & c_{23}c_{13}  
\end{array}
\right) \times U_{\mathrm{M}}
\end{equation}
is the PMNS (Pontecorvo-Maki-Nakagawa-Sakata) matrix. Here $c_{ij} = \cos \theta_{ij}$, $s_{ij} = \sin \theta_{ij}$, $\delta$ is the Dirac phase and $U_{\mathrm{M}} = \text{diag}(e^{i \varphi_1/2},e^{i \varphi_2/2},1)$ is a matrix containing the Majorana phases. In the following we will, however, neglect all the CP violating phases.

In order to determine the model parameters in terms of the quantities measured in neutrino oscillation experiments, the Yukawa matrix $y_{i\alpha}$ can be written using an adapted Casas-Ibarra parameterization~\cite{Casas:2001sr,Toma:2013zsa} as
\begin{equation}
\label{eq:casas-ibarra}
y=\sqrt{\Lambda}^{-1}R\sqrt{\hat{m}_\nu} \, U_{\mathrm{PMNS}}^{\dag} \, ,
\end{equation}
where $R$ is an complex orthogonal matrix, $R^T R=1$, that can be parameterized in terms of three angles  ($r_{1}$, $r_{2}$, $r_{3}$) in an analogous way to the neutrino mixing matrix --see equation (\ref{eq:PMNS}).  For simplicity, we take these three angles to be real so that the Yukawa couplings, $y_{i\alpha}$,  are real too. The general complex case would allow for $|R_{ij}| > 1$, implying larger Yukawa couplings. However, these scenarios involve a certain level of fine-tuning, in principle not preserved by radiative corrections \cite{AristizabalSierra:2011mn}.

The conservation of $\mathbb{Z}_2$ leads to the existence of a stable particle: the lightest particle charged under $\mathbb{Z}_2$. If neutral, it will constitute a good dark matter candidate. There are, therefore, two dark matter candidates in the scotogenic model: the lightest singlet fermion $N_1$ and the lightest neutral   $\eta$ scalar ($\eta_R$ or $\eta_I$).  Scenarios with scalar dark matter resemble the inert doublet model \cite{Barbieri:2006dq,LopezHonorez:2006gr,Honorez:2010re}, and have been studied in \cite{Hambye:2009pw,Kashiwase:2012xd,Klasen:2013jpa}. In this paper we will concentrate on $N_1$ dark matter~\cite{Kubo:2006yx, Sierra:2008wj, Suematsu:2009ww,Schmidt:2012yg}.
\section{Lepton flavor violating processes}
\label{sec:lfv}

The field of lepton flavor violation is about to begin a golden age. Several experimental projects will take place in the next few years, aiming at a discovery that could provide a valuable hint on new physics beyond the SM or, at least, at pushing the current bounds to much tighter values.

Currently, muon decay experiments provide the most stringent limits for most models. In their search for the muon radiative decay $\mu \to e \gamma$, the MEG collaboration has been able to set the impressive bound $\text{BR}(\mu \to e \gamma) < 5.7 \cdot 10^{-13}$ \cite{Adam:2013mnn}. This is expected to be improved to about $6 \cdot 10^{-14}$ after 3 years of acquisition time with the upgraded MEG II experiment \cite{Baldini:2013ke}. In what concerns the 3-body decay $\mu \to 3 \, e$, the future Mu3e experiment announces a sensitivity of $\sim 10^{-16}$ \cite{Blondel:2013ia}, which would imply a 4 orders of magnitude improvement on the current bound, $\text{BR}(\mu \to 3 \, e) <10^{-12}$, set long ago by the SINDRUM experiment \cite{Bellgardt:1987du}.

The LFV process where the most remarkable developments are expected is neutrinoless $\mec$ conversion in muonic atoms. In the near future, many competing experiments will search for a positive signal. These include Mu2e
\cite{Glenzinski:2010zz,Abrams:2012er}, DeeMe \cite{Aoki:2010zz,Natori:2014yba}, COMET \cite{Cui:2009zz,Kuno:2013mha} and PRISM/PRIME \cite{PRIME}. The expected sensitivities for the conversion rate range from $10^{-14}$ in the near future to an impressive $10^{-18}$ in the longer term, in all cases improving on previous experimental limits.

The current limits on $\tau$ observables are less stringent, but will also get improved in the near future by the LHCb collaboration \cite{Aaij:2014azz}, as well as by B-factories such as Belle II \cite{Bevan:2014iga}. In addition,  LFV can also be constrained by searches  at high-energy  colliders.  The CMS collaboration, for instance, recently reported the results of their search for $h \to \mu \tau$ \cite{CMS:2014hha}. A recent review of the status of the major experiments that will be soon searching for lepton flavor violation in charged lepton processes can be found in  \cite{Mihara:2013zna}.  For reference, in table~\ref{tab:sensi} we collect current bounds and expected near-future sensitivities for the most important low-energy LFV observables.

\begin{table}[tb!]
\centering
\begin{tabular}{|c|c|c|}
\hline
LFV Process & Present Bound & Future Sensitivity  \\
\hline
    $\mu \rightarrow  e \gamma$ & $5.7\times 10^{-13}$~\cite{Adam:2013mnn}  & $6\times 10^{-14}$~\cite{Baldini:2013ke} \\
    $\tau \to e \gamma$ & $3.3 \times 10^{-8}$~\cite{Aubert:2009ag}& $ \sim3\times10^{-9}$~\cite{Aushev:2010bq}\\
    $\tau \to \mu \gamma$ & $4.4 \times 10^{-8}$~\cite{Aubert:2009ag}& $ \sim3\times10^{-9}$~\cite{Aushev:2010bq} \\
    $\mu \rightarrow e e e$ &  $1.0 \times 10^{-12}$~\cite{Bellgardt:1987du} &  $\sim10^{-16}$~\cite{Blondel:2013ia} \\
    $\tau \rightarrow \mu \mu \mu$ & $2.1\times10^{-8}$~\cite{Hayasaka:2010np} & $\sim 10^{-9}$~\cite{Aushev:2010bq} \\
    $\tau^- \rightarrow e^- \mu^+ \mu^-$ &  $2.7\times10^{-8}$~\cite{Hayasaka:2010np} & $\sim 10^{-9}$~\cite{Aushev:2010bq} \\
    $\tau^- \rightarrow \mu^- e^+ e^-$ &  $1.8\times10^{-8}$~\cite{Hayasaka:2010np} & $\sim 10^{-9}$~\cite{Aushev:2010bq} \\
    $\tau \rightarrow e e e$ & $2.7\times10^{-8}$~\cite{Hayasaka:2010np} &  $\sim 10^{-9}$~\cite{Aushev:2010bq} \\
    $\mu^-, \mathrm{Ti} \rightarrow e^-, \mathrm{Ti}$ &  $4.3\times 10^{-12}$~\cite{Dohmen:1993mp} & $\sim10^{-18}$~\cite{PRIME} \\
    $\mu^-, \mathrm{Au} \rightarrow e^-, \mathrm{Au}$ & $7\times 10^{-13}$~\cite{Bertl:2006up} & \\
    $\mu^-, \mathrm{Al} \rightarrow e^-, \mathrm{Al}$ &  & $10^{-15}-10^{-18}$ \\
    $\mu^-, \mathrm{SiC} \rightarrow e^-, \mathrm{SiC}$ &  & $10^{-14}$~\cite{Natori:2014yba} \\
\hline
\end{tabular}
\caption{Current experimental bounds and future sensitivities for the most important LFV observables.}
\label{tab:sensi}
\end{table}

We now proceed to present analytical results for the LFV processes $\ell_\alpha\to\ell_\beta\gamma$, $\ell_\alpha \to 3 \, \ell_\beta$ and $\mec$
conversion in nuclei in the scotogenic model. For more details see  \cite{Toma:2013zsa}.

Let us first discuss radiative lepton decays. The branching fraction for $\ell_\alpha \to \ell_\beta \gamma$ is given by
\begin{equation}
\mathrm{BR}\left(\ell_{\alpha}\to\ell_{\beta}\gamma\right)=
\frac{3(4\pi)^3 \alpha_{\mathrm{em}}}{4G_F^2} 
|A_D|^2
\mathrm{Br}\left(\ell_{\alpha}\to\ell_{\beta}\nu_{\alpha}
\overline{\nu_{\beta}}\right) \, .
\end{equation}
Here $G_F$ is the Fermi constant and $\alpha_{\mathrm{em}}=e^2/(4\pi)$ is the electromagnetic fine structure constant, with $e$ the electromagnetic
coupling. $A_D$ is the dipole form factor, given by
\begin{equation}
A_D = \sum_{i=1}^3\frac{y_{i\beta}^*y_{i\alpha}}
{2(4\pi)^2}\frac{1}{m_{\eta^+}^2}
F_2\left(\xi_i\right) \, ,
\label{eq:AD}
\end{equation}
where the $\xi_i$ parameters are defined as $\xi_i\equiv M_{N_i}^2/m_{\eta^+}^2$ and the loop function $F_2(x)$ is given in appendix \ref{sec:appendix1}.

We now turn to the 3-body decays $\ell_\alpha \to 3 \, \ell_\beta$. The branching ratio is given by a slightly more involved expression
\begin{eqnarray}
\text{BR}\left(\ell_{\alpha}\to
\ell_{\beta}\overline{\ell_{\beta}}\ell_{\beta}\right)&=&
\frac{3(4\pi)^2\alpha_{\mathrm{em}}^2}{8G_F^2}
\left[|A_{ND}|^2
  +|A_D|^2\left(\frac{16}{3}\log\left(\frac{m_\alpha}{m_\beta}\right)
  -\frac{22}{3}\right)\right.\nonumber\\
 &&\left.+\frac{1}{6}|B|^2+\left(-2 A_{ND} A_D^{*}+\frac{1}{3}A_{ND} B^*
  -\frac{2}{3}A_D B^*+\mathrm{h.c.}\right)\right]\nonumber\\
&&\times \, \mathrm{Br}\left(\ell_{\alpha}\to\ell_{\beta}\nu_{\alpha}
\overline{\nu_{\beta}}\right) \, . \label{eq:l3lBR}
\end{eqnarray}
Here we have kept $m_\beta \ll m_\alpha$ only in the logarithmic term, where it avoids the appearance of an infrared divergence. The form factor $A_D$ is generated by dipole photon penguins and is given in equation \eqref{eq:AD}. Regarding the other form factors, $A_{ND}$, given by
\begin{equation}
A_{ND}=\sum_{i=1}^3\frac{y_{i\beta}^*y_{i\alpha}}
{6(4\pi)^2}\frac{1}{m_{\eta^+}^2}
G_2\left(\xi_i\right), \label{eq:AND}
\end{equation}
is generated by non-dipole photon penguins, whereas $B$, induced by box diagrams, is 
\begin{equation}
e^2 B = \frac{1}{(4\pi)^2m_{\eta^+}^2} 
 \sum_{i,\:j=1}^3\left[ \frac{1}{2} D_1(\xi_i,\xi_j) y_{j \beta}^* y_{j \beta}
	   y_{i \beta}^* y_{i \alpha} + \sqrt{\xi_i\xi_j}
	   D_2(\xi_i,\xi_j) y_{j \beta}^* y_{j \beta}^* y_{i \beta}
	   y_{i \alpha}  \right] \, .
\label{eq:B}
\end{equation}
The loops functions $G_2(x)$, $D_1(x,y)$ and $D_2(x,y)$ are defined in Appendix \ref{sec:appendix1}.

We note that the $Z$-boson penguin contributions are negligible, since in this model they are suppressed by charged lepton masses~\cite{Toma:2013zsa}. Similarly suppressed are the  Higgs-penguin contributions, which we have not included in our calculations. 

Next, we consider $\mec$ conversion in nuclei. This is the LFV process with the most remarkable experimental projects in the next few years.  The conversion rate, normalized to the the muon capture rate, can be expressed as~\cite{Kuno:1999jp,Arganda:2007jw}
\begin{align}
{\rm CR} (\mec, {\rm Nucleus}) &= 
\frac{p_e \, E_e \, m_\mu^3 \, G_F^2 \, \alpha_{\mathrm{em}}^3 
\, Z_{\rm eff}^4 \, F_p^2}{8 \, \pi^2 \, Z \, \Gamma_{\rm capt}}  \nonumber \\
&\times \left\{ \left| (Z + N) \left( g_{LV}^{(0)} + g_{LS}^{(0)} \right) + 
(Z - N) \left( g_{LV}^{(1)} + g_{LS}^{(1)} \right) \right|^2 + 
\right. \nonumber \\
& \ \ \ 
 \ \left. \,\, \left| (Z + N) \left( g_{RV}^{(0)} + g_{RS}^{(0)} \right) + 
(Z - N) \left( g_{RV}^{(1)} + g_{RS}^{(1)} \right) \right|^2 \right\} \,.
\end{align}
Here $Z$ and $N$ are the number of protons and neutrons in the nucleus, $Z_{\rm eff}$ is the effective atomic charge \cite{Chiang:1993xz}, $F_p$ denotes the nuclear matrix element and $\Gamma_{\rm capt}$ represents the total muon capture rate. The values of these parameters depend on the considered nucleus. For the nuclei used in current or near future experiments, these values can be found in
\cite{Arganda:2007jw} and references therein. Furthermore, $p_e$ and $E_e$ ($\simeq m_\mu$ in the numerical evaluation) are the momentum and energy of the electron. In the above, $g_{XK}^{(0)}$ and $g_{XK}^{(1)}$ (with $X = L, R$ and $K = S, V$) are generally given by
\begin{align}
g_{XK}^{(0)} &= \frac{1}{2} \sum_{q = u,d,s} \left( g_{XK(q)} G_K^{(q,p)} +
g_{XK(q)} G_K^{(q,n)} \right)\,, \nonumber \\
g_{XK}^{(1)} &= \frac{1}{2} \sum_{q = u,d,s} \left( g_{XK(q)} G_K^{(q,p)} - 
g_{XK(q)} G_K^{(q,n)} \right)\,.
\end{align}
The numerical values of the $G_K$ coefficients can be found in \cite{Kuno:1999jp,Kosmas:2001mv,Arganda:2007jw}. In the scotogenic model, the effective couplings $g_{XK(q)}$ receive several contributions, and thus they can be split as
\begin{eqnarray}
g_{LV(q)} &\approx& g_{LV(q)}^{\gamma} \,, \nonumber \\
g_{RV(q)} &=& \left. g_{LV(q)} \right|_{L \leftrightarrow R}\,, \nonumber \\
g_{LS(q)} &\approx& 0 \, , \nonumber \\ 
g_{RS(q)} &\approx& 0 \, ,
\end{eqnarray}
where $g_{LV(q)}^{\gamma}$ is generated by photon penguins. We note that in the scotonic model there are no box contributions to $\mec$ conversion in nuclei (besides the negligible SM contribution) due to the $\mathbb{Z}_2$ symmetry, which forbids the coupling between the $\eta^\pm$ scalars and the quark sector. Regarding the $Z$-boson penguins contributions, they turn out to be suppressed by charged lepton masses, see \cite{Toma:2013zsa} for more details. The $g_{LV(q)}^{\gamma}$ effective coupling can be written as
\begin{equation}
g_{LV(q)}^{\gamma} = \frac{\sqrt{2}}{G_F} e^2 Q_q 
\left(A_{ND} - A_D \right)\,. 
\end{equation}
The form factors $A_{ND}$ and $A_D$ have been already defined in Eqs. \eqref{eq:AND} and \eqref{eq:AD}. Furthermore, $Q_q$ is the electric charge of the corresponding quark.
\section{Results}
\label{sec:res}

In this section, we assess the impact of future LFV experiments on the scotogenic model by means of a random scan over its entire parameter space. That is, we first find a large sample of models compatible with current data--in particular neutrino masses, $\meg$ and dark matter-- and then study what region of this viable parameter space will be probed by future experiments.

In our random scan, we take as free parameters of the model the following:
\begin{equation}
M_{N_i}, m_R, m_{\eta^+},\lambda_5, r_{1},r_{2},r_{3}.
\end{equation}
From them,  one can reconstruct the original Lagrangian parameters, in particular the Yukawa couplings, $y_{i\alpha}$, using equations \eqref{eq:mass1} \eqref{eq:mass2}, \eqref{eq:mass3} and \eqref{eq:casas-ibarra}.  All our models are, by construction, automatically consistent with current neutrino data (at $3\sigma$) \cite{Capozzi:2013csa}.

These free parameters are subject to a number of theoretical and experimental constraints, which we now describe. First of all, we impose a perturbativity limit on the Yukawa and scalar couplings: $|y_{i\alpha}|,|\lambda_j|<3$. The scalar couplings, $\lambda_j$, are further required to satisfy the vacuum stability conditions and we ensure compatibility with electroweak precision tests \cite{Barbieri:2006dq,Goudelis:2013uca}. 

Direct searches at colliders impose a lower bound on the masses of the scalar particles \cite{Pierce:2007ut,Lundstrom:2008ai} . For definiteness, we require all scalars to be above $100\GeV$. To ensure that the scan is sufficiently general, the upper bound on the scalar masses was set at $5\TeV$. The masses of the singlet fermions, $m_{N_i}$, are not constrained by current collider data. In the scan, we allowed them to vary from a common minimum value of $1\GeV$ to a maximum value of $3.3$, $5$ and $10\TeV$ respectively for $i=1,2,3$. 

The dark matter particle in this model can be a neutral scalar ($\eta_{R,I}$) or a singlet fermion ($N_1$). Both possibilities have been examined in the previous literature and it is known that they give rise to a different phenomenology. We assume in the following that the dark matter is the singlet fermion and that its relic density is the result of a freeze-out process in the early Universe (freeze-in~\cite{Hall:2009bx,Molinaro:2014lfa} is an alternative possibility we do not consider), as this is the most interesting scenario from the point of view of LFV processes.  In this case, the dark matter relic density is determined by the $N_1$ annihilation rate, which  depends on the Yukawa couplings. Since they must  be large enough to explain the observed dark matter density, the rates of LFV processes, which are proportional to these Yukawas, are generally expected to be observable. We examine two different dark matter scenarios  depending on the process that sets the value of the relic density: $N_1$-$N_1$ annihilations or  $N_1$-$\eta$ coannihilations ($N_1$-$N_2$ coannihilations are rarely relevant as they depend on the same Yukawa couplings as $N_1$-$N_1$). For each case, we require the corresponding process to be dominant. We have implemented the scotogenic model into micrOMEGAs~\cite{Belanger:2013oya}, which accurately computes the relic density taking into account all relevant effects, including resonances and coannihilations. All our viable models are consistent with the observed value of the dark matter density, as determined by WMAP~\cite{Hinshaw:2012aka} and PLANCK~\cite{Ade:2013zuv}.

Regarding LFV processes, we have only imposed the current bound on $\meg$, which is usually assumed to set the strongest constraint. In this way, we can actually test this assumption by  comparing the rates of other LFV processes against their current experimental limits.

Models satisfying the above mentioned constraints are  called \emph{viable} models in the following. We have generated a sample of about $10^5$ of them for each of the two dark matter scenarios, which we will discuss separately. Our analysis is based on this sample of viable models.

\subsection{Dark matter via $N_1$-$N_1$ annihilations}
\begin{figure}[t]
\begin{center} 
\includegraphics[scale=0.45]{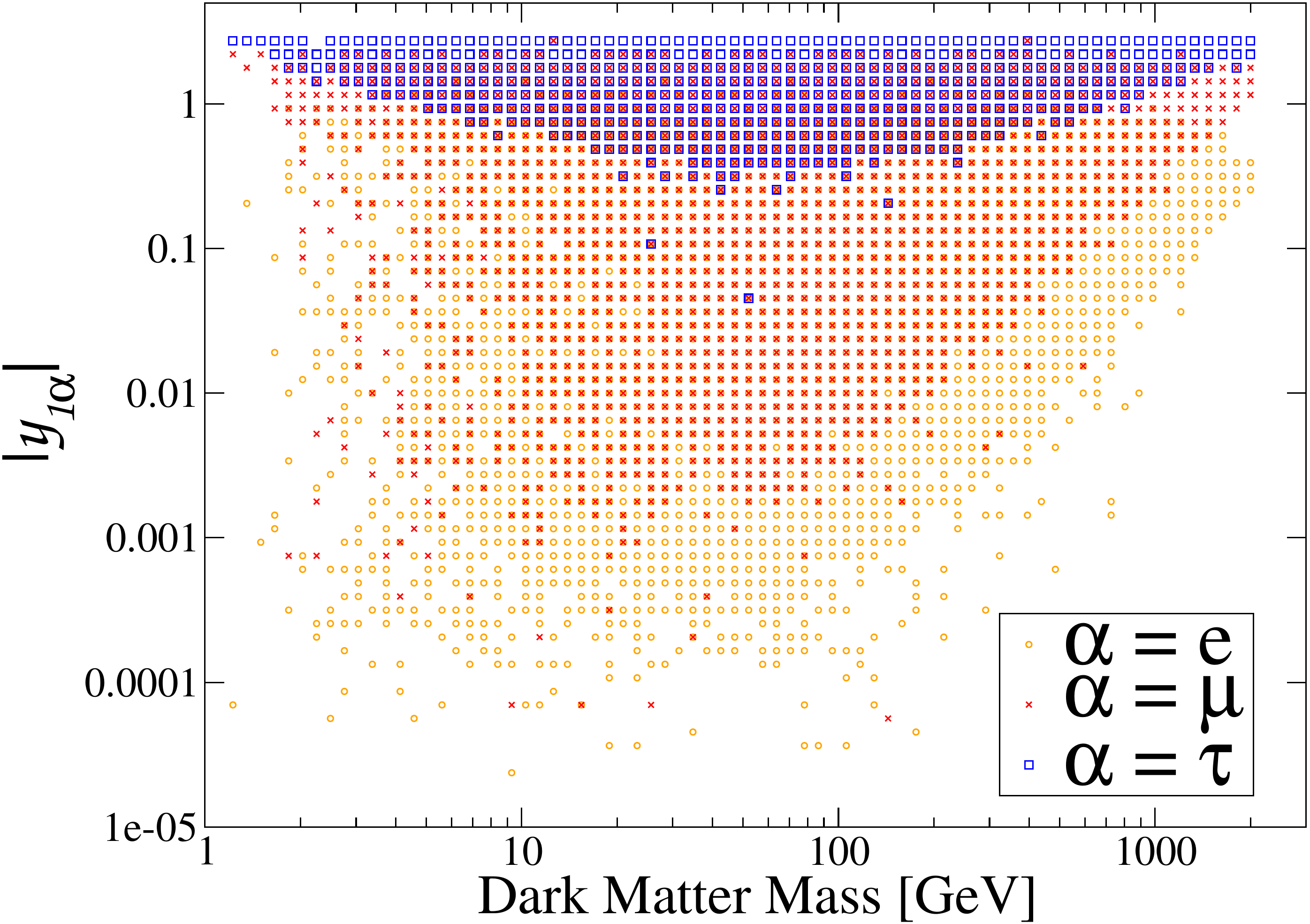}
\caption{\small The Yukawa couplings associated with the dark matter particle, $y_{1i}$, as a function of $\mdm$. \label{fig:nocoanndmyuk}}
\end{center}
\end{figure}

To begin with, let us consider viable models where the relic density is   determined by annihilation processes only, without coannihilation effects. This is the most favorable case for LFV processes because the Yukawa couplings tend to be rather large. First, we will describe the resulting parameter space and then study the prospects for a positive observation in future LFV experiments. As we will show, this scenario can be entirely probed by such experiments. 

\begin{figure}[t]
\begin{center} 
\includegraphics[scale=0.45]{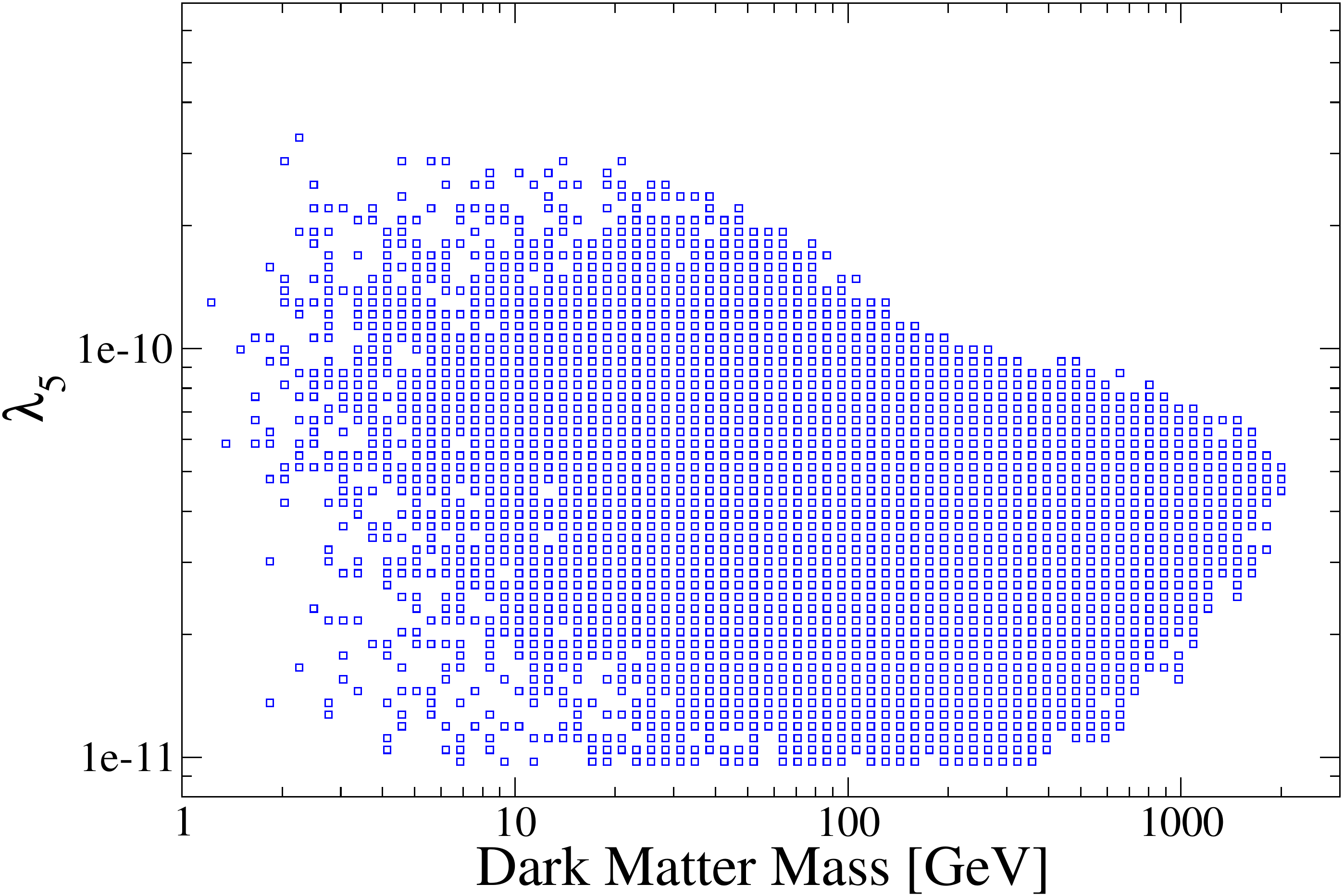}
\caption{\small The allowed values of $\lambda_5$ as a function of $\mdm$. \label{fig:nocoannla5}}
\end{center}
\end{figure}

The dark matter annihilation cross section is determined by the Yukawa couplings $y_{1\alpha}$. At least one of these couplings must, therefore, be large enough to ensure that the dark matter density is consistent with the observations.  Figure \ref{fig:nocoanndmyuk} shows the viable models in the plane ($\mdm$, $\left|y_{1\alpha}\right|$) for $\alpha=e$ (yellow circles), $\mu$ (red crosses), and $\tau$ (blue squares). They feature  masses between $1\GeV$ and $2\TeV$, and dark matter Yukawa couplings between $10^{-5}$ and $3$ (our perturbativity limit). As a result of the $\meg$ bound, these couplings satisfy the hierarchy $\left|y_{1e}\right|\lesssim \left|y_{1\mu}\right|\lesssim \left|y_{1\tau}\right|$, with $\left|y_{1\tau}\right|$ rarely lying below $0.3$ and $\left|y_{1e}\right|$ rarely going above that value.   Thus, dark matter annihilates mainly into third-family leptons: $\tau^+\tau^-$ and $\bar\nu_\tau \nu_\tau$.  Since the perturbativity limit on $y_{1\tau}$ is saturated at both the lowest and the highest dark matter masses, we can claim that it is not possible to find viable models outside this mass range. Notice also that this range is quite sensitive to the exact value of the perturbativity limit imposed. Had we used the more restrictive condition $|y_{i\alpha}|<1$, the lowest and highest value for the dark matter mass would have changed to about $10\GeV$ and $700\GeV$ respectively.

The allowed values of $\lambda_5$ are shown in figure \ref{fig:nocoannla5}. They turn out to be restricted to the range ($10^{-11}$, $4\times 10^{-10}$), with most points lying close to $10^{-10}$. $\lambda_5$ indeed has to be very small in this setup. This small value of $\lambda_5$ and its narrow range of variation are the result of the interplay between the perturbativity limit, neutrino masses and the dark matter constraint.  Due to neutrino masses, larger values of $\lambda_5$  generically imply smaller Yukawa couplings, which would lead to a dark matter density larger than the observed one; smaller values of $\lambda_5$, instead, tend to be give Yukawa couplings above the perturbativity limit.  Let us  emphasize that $\lambda_5$ is naturally small in the sense of 't Hooft~\cite{'tHooft:1979bh}, because in the limit $\lambda_5 \to 0$ lepton number is restored~\cite{Kubo:2006yx}. Furthermore, we also note that the allowed values of $\lambda_5$ found in our analysis depend on our assumption of a real $R$ matrix.

\begin{figure}[t]
\begin{center} 
\includegraphics[scale=0.45]{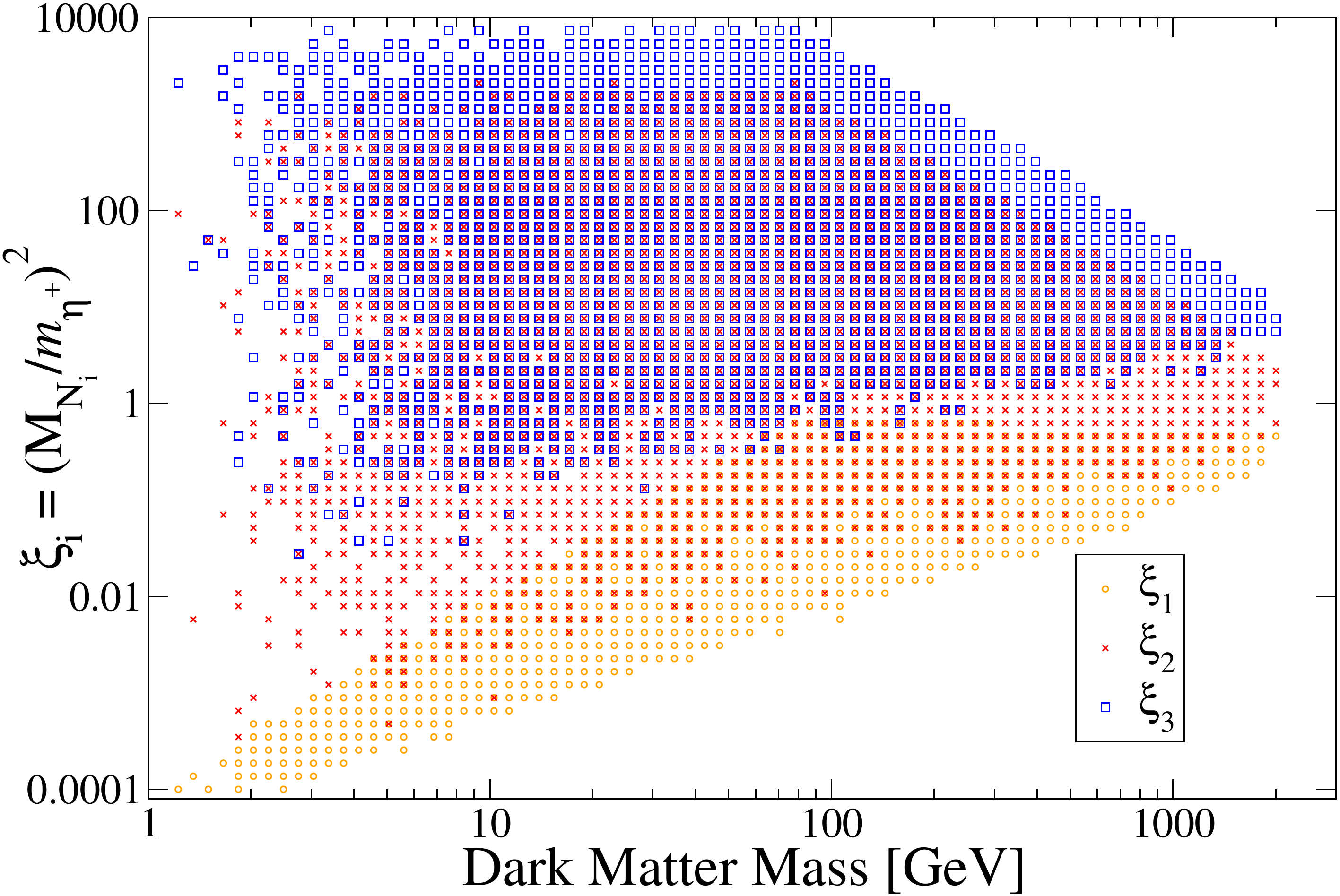}
\caption{\small The parameters $\xi_{i}=(M_{N_i}/m_{\eta^+})^2$ as a function of $\mdm$. \label{fig:nocoannxi}}
\end{center}
\end{figure}

\begin{figure}[t]
\begin{tabular}{cc}
\includegraphics[scale=0.23]{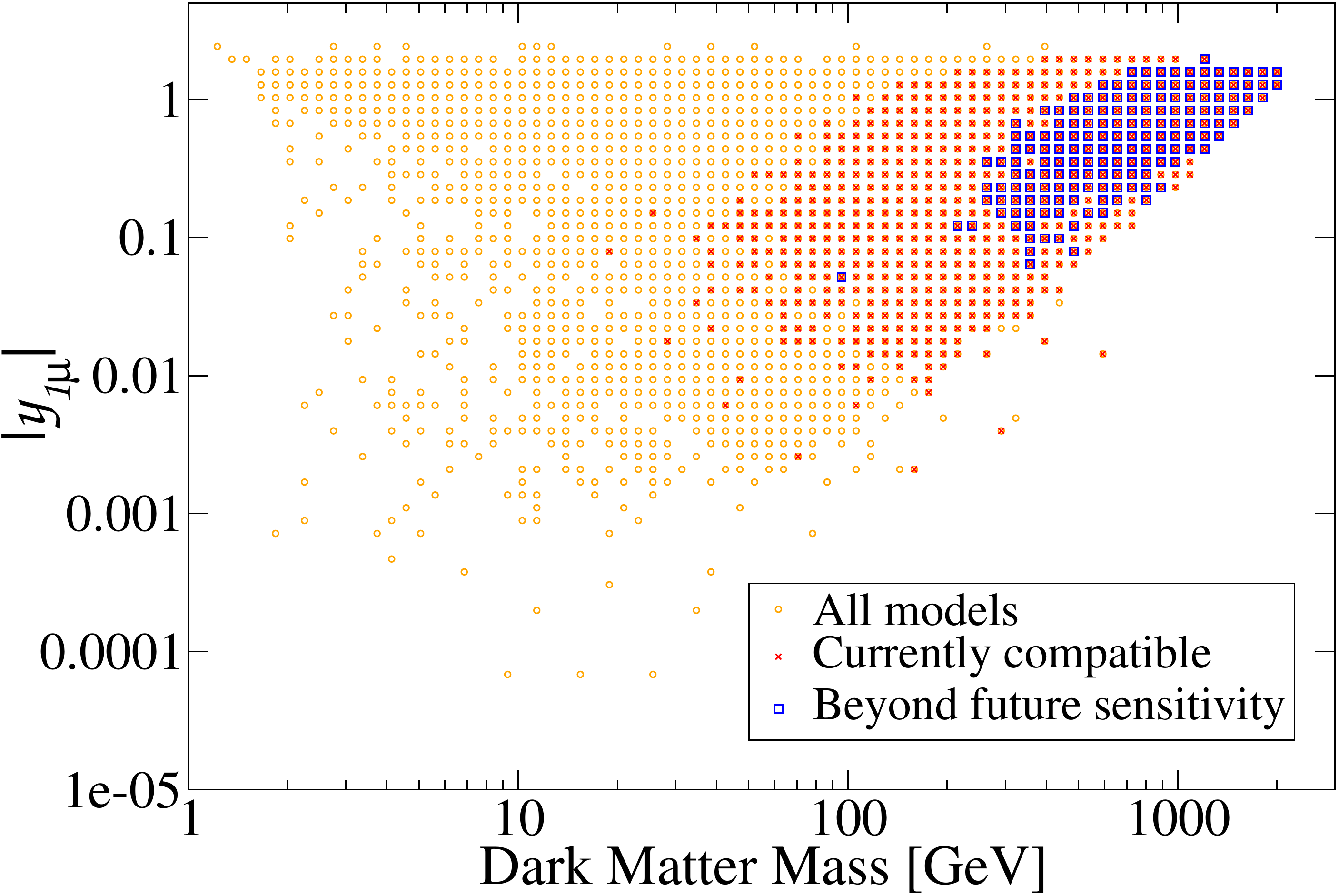} &    \includegraphics[scale=0.23]{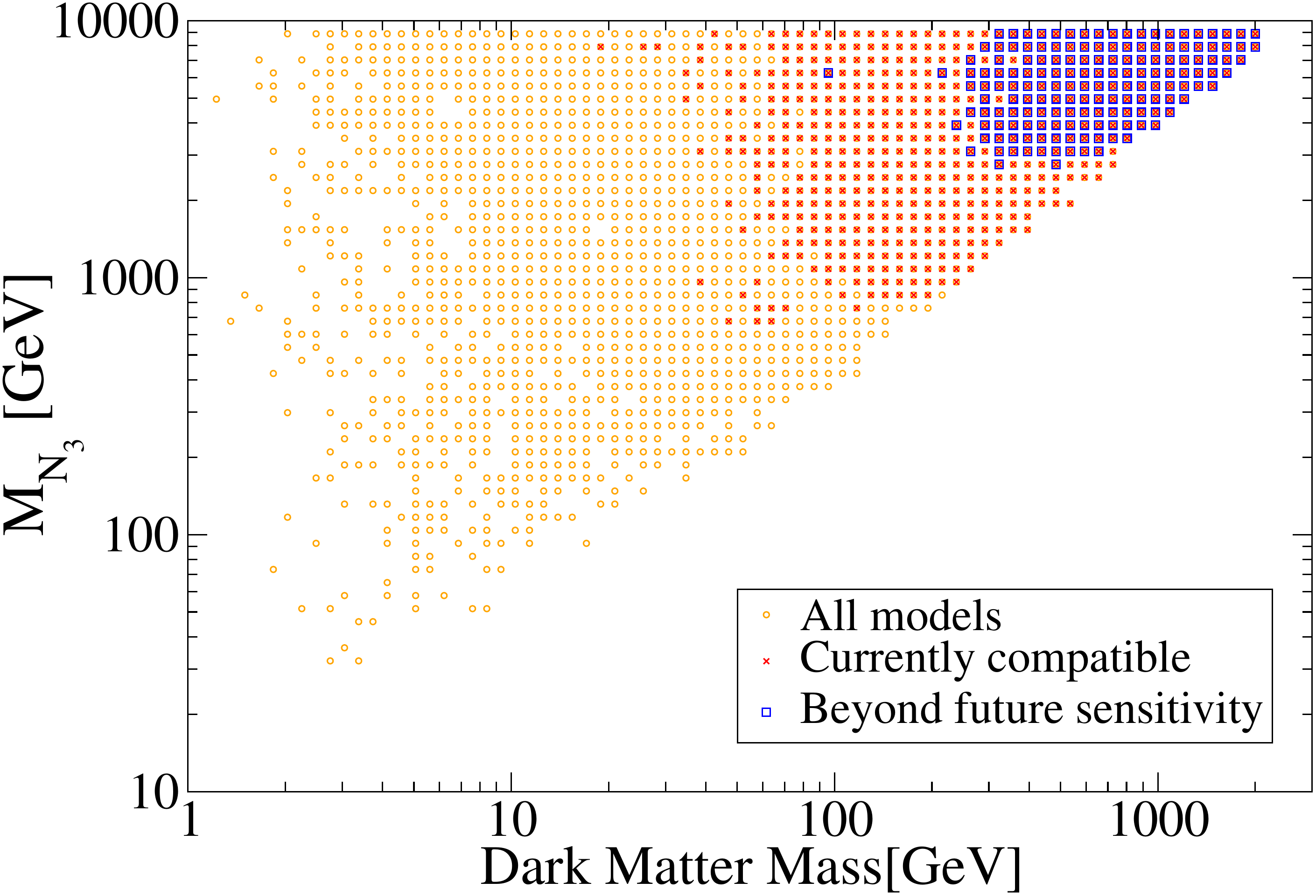} \\
(a) &    (b)
\end{tabular}
\caption{\small (a) A scatter plot of $\left|y_{1\mu}\right|$ versus $\mdm$ for all the models in our sample (yellow circles), those compatible with current bounds on LFV $\tau$ decays (red crosses), and those that are beyond the expected sensitivity of future searches for LFV $\tau$ decays (blue squares). (b) Similar to  (a) but for $M_{N_3}$ versus $\mdm$.  \label{fig:nocoanntaufuture}}
\end{figure}

As shown in the previous section, the rates for the different LFV processes depend on loop functions of the parameters $\xi_i=m_{N_i}^2/m_{\eta^+}^2$. Figure \ref{fig:nocoannxi} shows a scatter plot of the dark matter mass versus $\xi_i$. Since $N_1$ is the lightest odd particle, $\xi_1$ is always smaller than $1$, reaching values of order $10^{-4}$ for $\mdm\sim \mathrm{few}\GeV$. In principle smaller values of $\xi_1$ are possible in our scan, but they are not realized within the viable models. They always feature, for instance, a light charged scalar  ($\mch\lesssim 300\GeV$) and small dark matter masses,  $\mdm\lesssim 10\GeV$.  $\xi_2$ and $\xi_3$, on the other hand, can be either larger or smaller than $1$. We see that $\xi_3$ tends to be larger than $1$,  reaching values as high as $10^4$. Interestingly, we have found that viable models never feature a degenerate or quasi-degenerate spectrum for the singlet fermions ($\xi_1\approx\xi_2\approx\xi_3$). The ratio $M_{N_3}/M_{N_1}$, in fact, has a minimum value of about $3.5$ in our sample --see also the right-hand side of figure \ref{fig:nocoanntaufuture}.

\begin{figure}[t]
\begin{tabular}{cc}
\includegraphics[scale=0.23]{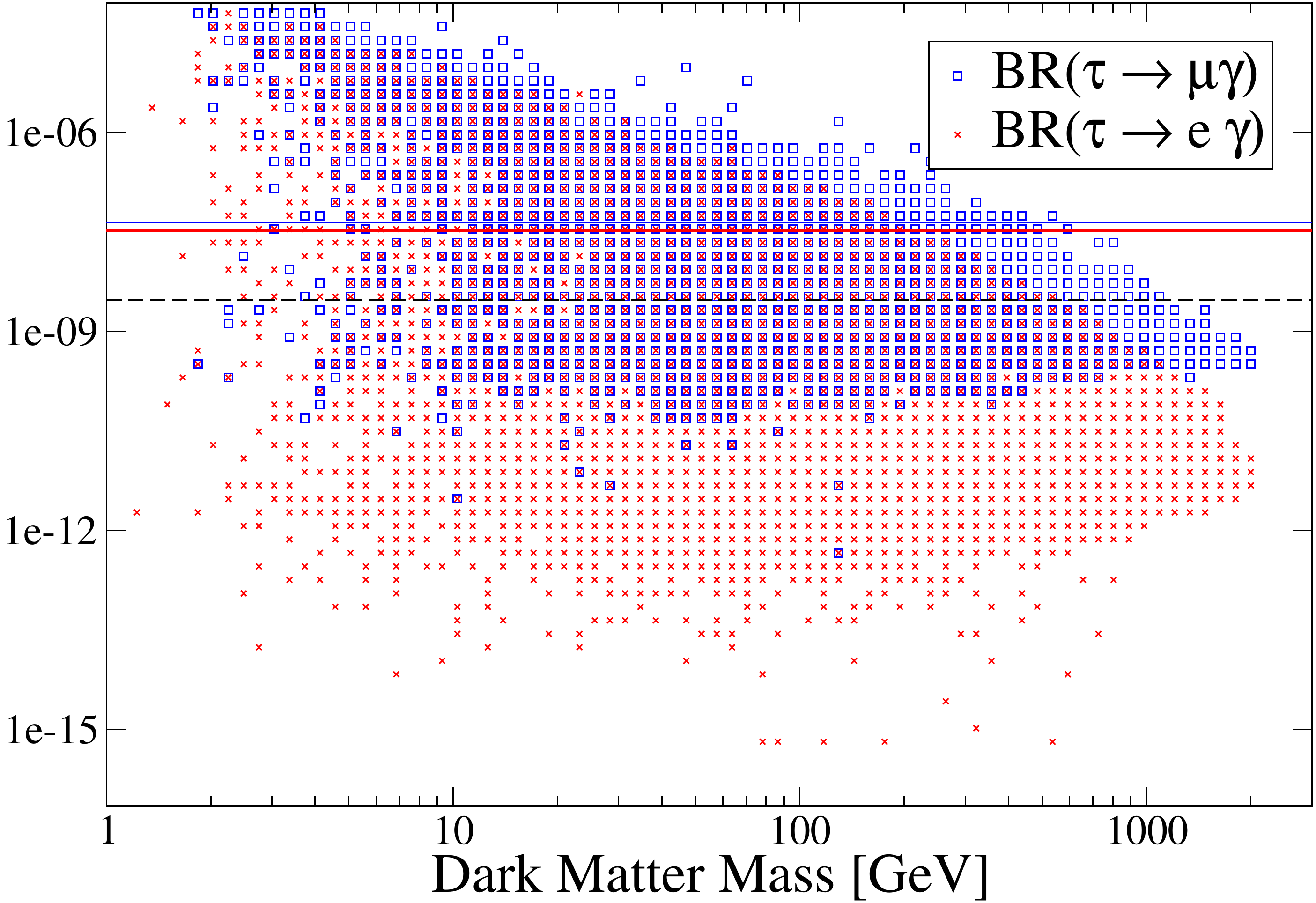} &    \includegraphics[scale=0.23]{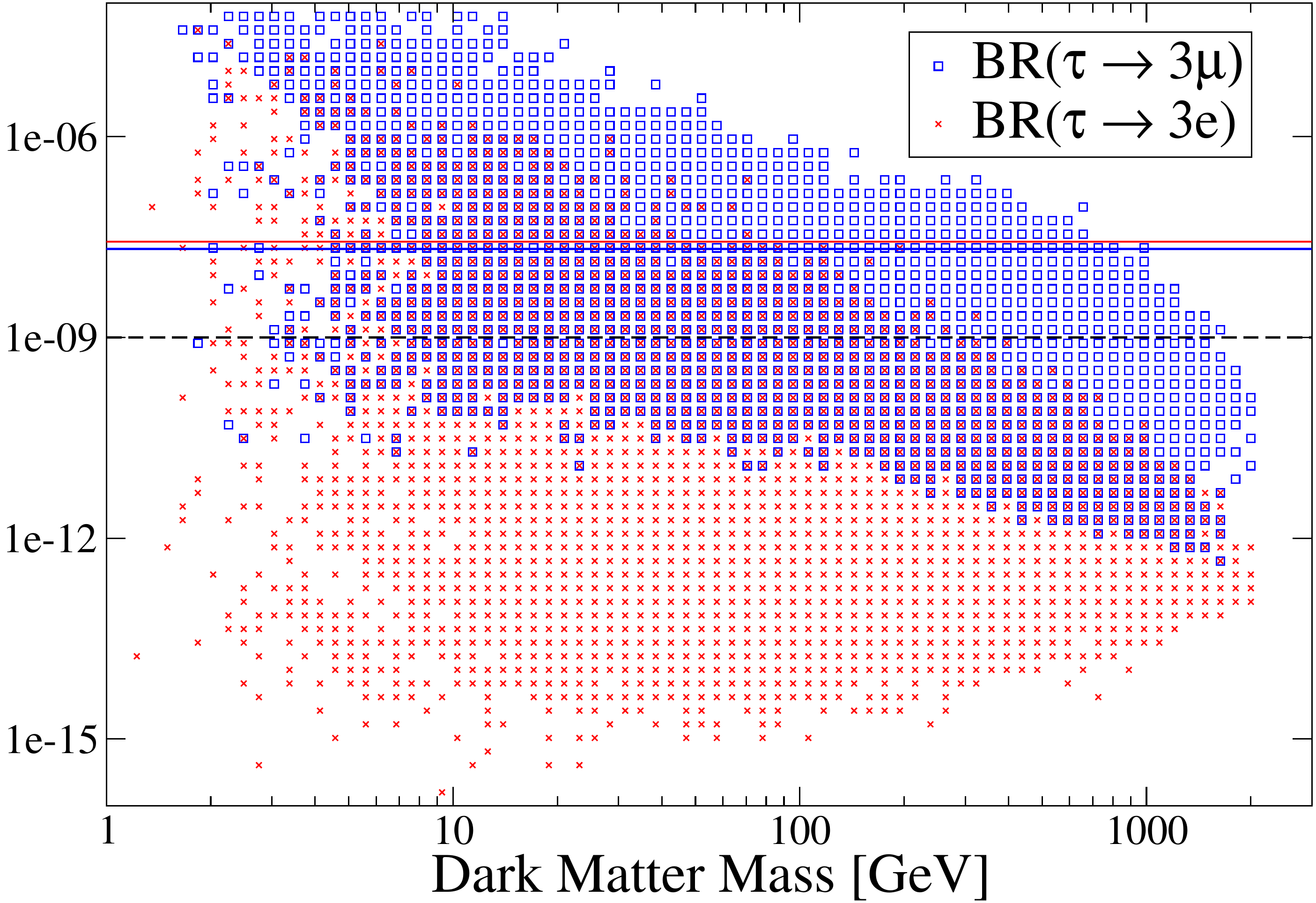} \\
(a) &    (b)
\end{tabular}
\caption{\small (a) The rates of $\tau\to \mu\gamma$ (blue squares) and $\tau\to  e\gamma$ (red crosses) as a function of the dark matter mass. The current bounds for both processes are shown with solid blue and red lines, respectively.  The expected future sensitive for both processes is also displayed as a dashed black line. (b) Similar to the figure in (a) but for the processes $\tau\to 3\mu$ (blue squares) and $\tau\to3e$ (red crosses).\label{fig:nocoanntaulg}}
\end{figure}

The branching ratios for the most important lepton flavor violating $\tau$ decays are shown in Figure \ref{fig:nocoanntaulg}.  Its left panel displays BR$(\tau\to \mu\gamma)$ (blue squares) and BR$(\tau\to e\gamma)$ (red crosses) versus the dark matter mass.  The current experimental bounds on these decays are also shown as solid lines, blue and red, respectively. We see  that the current bounds can be violated, particularly at low $\mdm$. Thus, in certain regions of the parameter space, $\tau\to \mu\gamma$ is more constraining than $\meg$, even if the former has a less stringent experimental bound. In any case, current bounds do not exclude the low $\mdm$ region, as one can also find models with smaller branching ratios there. Notice that $\tau\to\mu\gamma$ tends to have a branching ratio larger than $\tau\to e \gamma$, with most points featuring values above  $10^{-11}$ for BR$(\tau\to\mu\gamma)$ and above $10^{-14}$ for BR$(\tau\to e\gamma)$.  Planned experiments are expected to reach sensitivities of order $3\times 10^{-9}$ (dashed black line) for both of these decays --see table \ref{tab:sensi}. Even though significant, such improvement would not be sufficient to exclude this scenario or restrict the value of $\mdm$.

The right panel of Figure \ref{fig:nocoanntaulg} shows instead the branching ratios for the processes $\tau\to 3\mu$ (blue squares) and $\tau\to 3e$ (red crosses) versus the dark matter mass. The other conventions are the same as in the left panel. BR$(\tau\to 3\mu)$ varies approximately between $10^{-4}$ and $10^{-12}$ whereas BR$(\tau\to 3\mu)$ varies between $10^{-4}$ and $10^{-16}$.  The current experimental limits on these decays can therefore be violated at low $\mdm$, particularly for $\tau\to 3\mu$. The expected future sensitivity for both branching ratios  is of order  $10^{-9}$ (the dashed black line), and can  probe a large number of models. It is not enough, however, to cover the entire parameter space, as there are models featuring smaller branching ratios over the whole range of $\mdm$.

As we have seen, searches for LFV $\tau$ decays play an important role in this model. They can already exclude a large number of points that are compatible with the $\meg$ constraint, and could, in the future,  probe a significant region of the parameter space. In figure \ref{fig:nocoanntaufuture} we further illustrate these facts.  They differentiate  our entire set of viable points (yellow circles) from those that are compatible with all current bounds on LFV $\tau$ decays (red crosses), and from those that lie beyond the expected sensitivity of \emph{all} future searches for LFV $\tau$ decays (blue squares). Specifically,  these points are projected onto the planes $(\mdm,|y_{1\mu}|)$ in the left panel and $(\mdm,M_{N_3})$ in the right panel. From the left panel we learn that current bounds exclude the region of low $\mdm$ and large $\left|y_{1\mu }\right|$. In fact, no points compatible with the current limits on LFV  $\tau$ decays are found for $\mdm\lesssim 20\GeV$ and $\left|y_{1\mu }\right|\gtrsim 0.1$. Moreover, all the points giving branching ratios below  the expected sensitivity of future LFV $\tau$ searches are found  at large values of $\mdm$ or small values of $\left|y_{1\mu}\right|$. From the right panel we see that all points that cannot be probed by future experiments feature $M_{N_3}\gtrsim 3\TeV$ and $\mdm\gtrsim 100\GeV$. That is, they are characterized by a hierarchical spectrum of singlet fermions containing at least a very heavy particle. Thus, if the spectrum of singlet fermions were such that they all had masses below $3\TeV$, the entire parameter space of this model could be probed by  future  LFV experiments involving  $\tau$ decays only, without any additional information from $\mec$ processes.

\begin{figure}[t]
\begin{center} 
\includegraphics[scale=0.45]{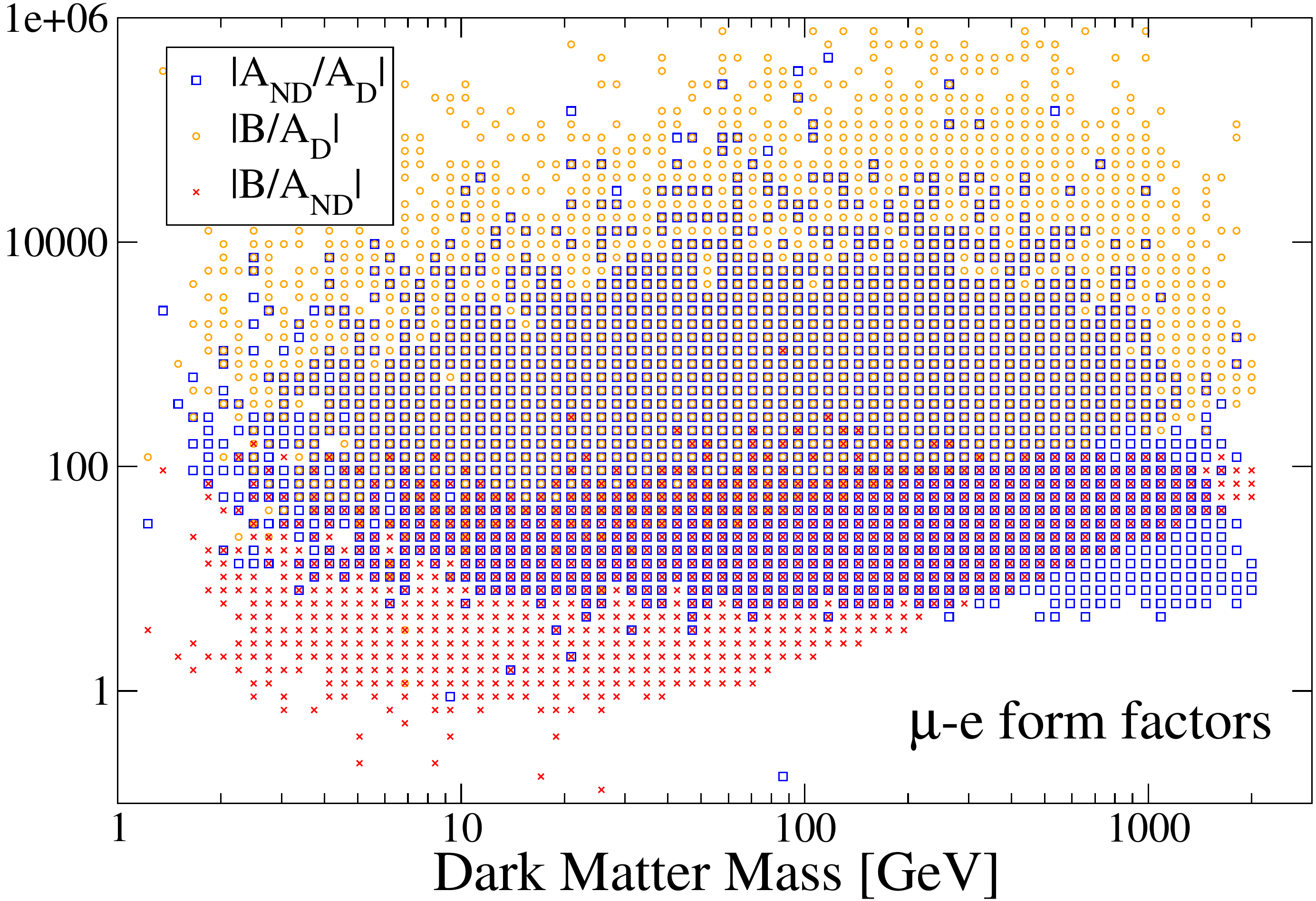}
\caption{\small The ratios between the different amplitudes that contribute to $\mec$ transitions as a function of the dark matter mass. \label{fig:nocoannampratio}}
\end{center}
\end{figure}

Let us now turn our attention to LFV processes in the $\mec$ sector. As explained in section \ref{sec:lfv}, all $\mec$ transitions in this model are determined by just three different form factors: the dipole ($A_D$), the non-dipole ($A_{ND}$), and the box ($B$). $\meg$ depends on $A_D$, $\mec$ conversion on the difference $A_D-A_{ND}$, and $\mte$ on  all three in a more complicated way. It is important to identify, therefore, which of these contributions dominates. Due to the strong bound on $\meg$, we expect $A_D$ to be suppressed with respect to the other two. And that is exactly what we find, as illustrated in figure \ref{fig:nocoannampratio}. It shows the ratio between the three amplitudes: $\left|A_{ND}/A_D\right|$ (blue squares), $\left|B/A_D\right|$ (orange circles) and $\left|B/A_{ND}\right|$ (red crosses). In most models we see in fact that  $A_{ND}>A_{D}$ and $B>A_D$, with their ratios reaching values as high as $10^{6}$. As a result, most points in our results have $\mte$ dominated by non-dipole and box contributions. In \cite{Toma:2013zsa}, it was stated instead that the non-dipole contribution never exceeds the dipole one. The reason for this discrepancy with our findings is  that such conclusion was reached for a scenario where the singlet fermions are degenerate, leading to the different result.
If the singlet fermion spectrum is not degenerate, as expected in general, one can indeed have $A_{ND}>A_D$. In our sample of viable models, not only is $A_{ND}\gg A_{D}$ possible but it is also the most likely result. Regarding the relation between $A_{ND}$ and $B$, we see that $B>A_{ND}$ in most models, $B$ being two orders of magnitude larger than $A_{ND}$ in some cases.

\begin{figure}[t]
\begin{center} 
\includegraphics[scale=0.45]{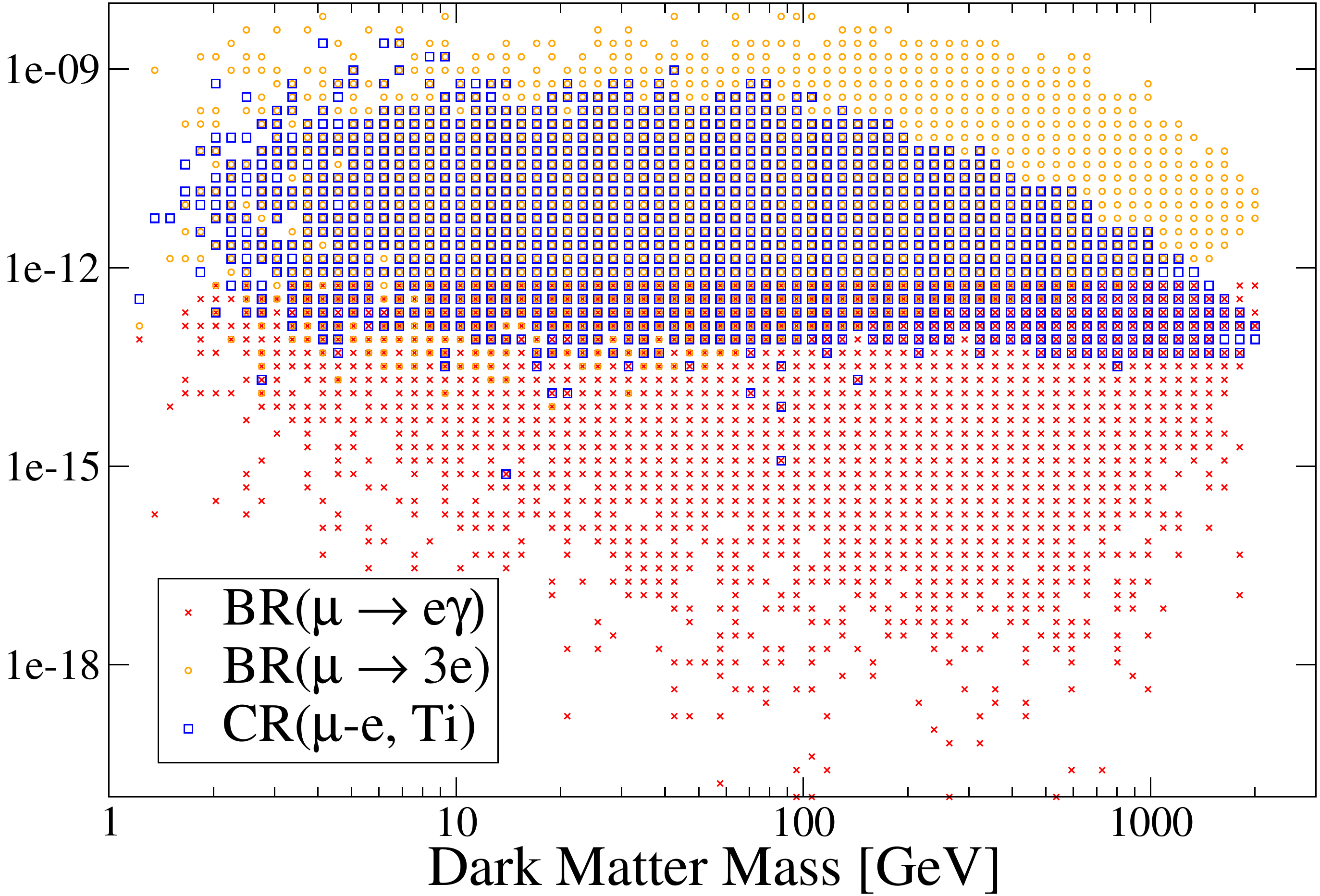}
\caption{\small The rates of $\meg$ (red crosses), $\mte$ (orange circles), and $\mec$ conversion in Titanium (blue squares) as a function of the dark matter mass.  \label{fig:nocoannalllfv}}
\end{center}
\end{figure}

Figure \ref{fig:nocoannalllfv} displays the  rates for the LFV processes involving $\mu-e$ transitions: BR($\meg$) (red crosses), BR($\mte$) (orange circles), and CR($\mec$, Ti) (blue squares). BR($\meg$) varies between its current experimental bound, which we imposed, and about $10^{-20}$. Current bounds on BR($\mte$) and CR($\mec$,Ti) were not imposed, and from the figure we see  that they can in fact be violated. Thus, $\meg$ alone does not guarantee compatibility with the current bounds on other $\mu$-$e$ lepton flavor violating processes. It is clear, nevertheless, that models compatible with all current bounds can be found over the entire range of the dark matter mass. Notice that most models feature BR$(\mte)>10^{-12}$  and CR$(\mec,\mathrm{Ti})>10^{-13}$, so they are never much below the current experimental bounds. 

\begin{figure}[t]
\begin{tabular}{cc}
 \includegraphics[scale=0.23]{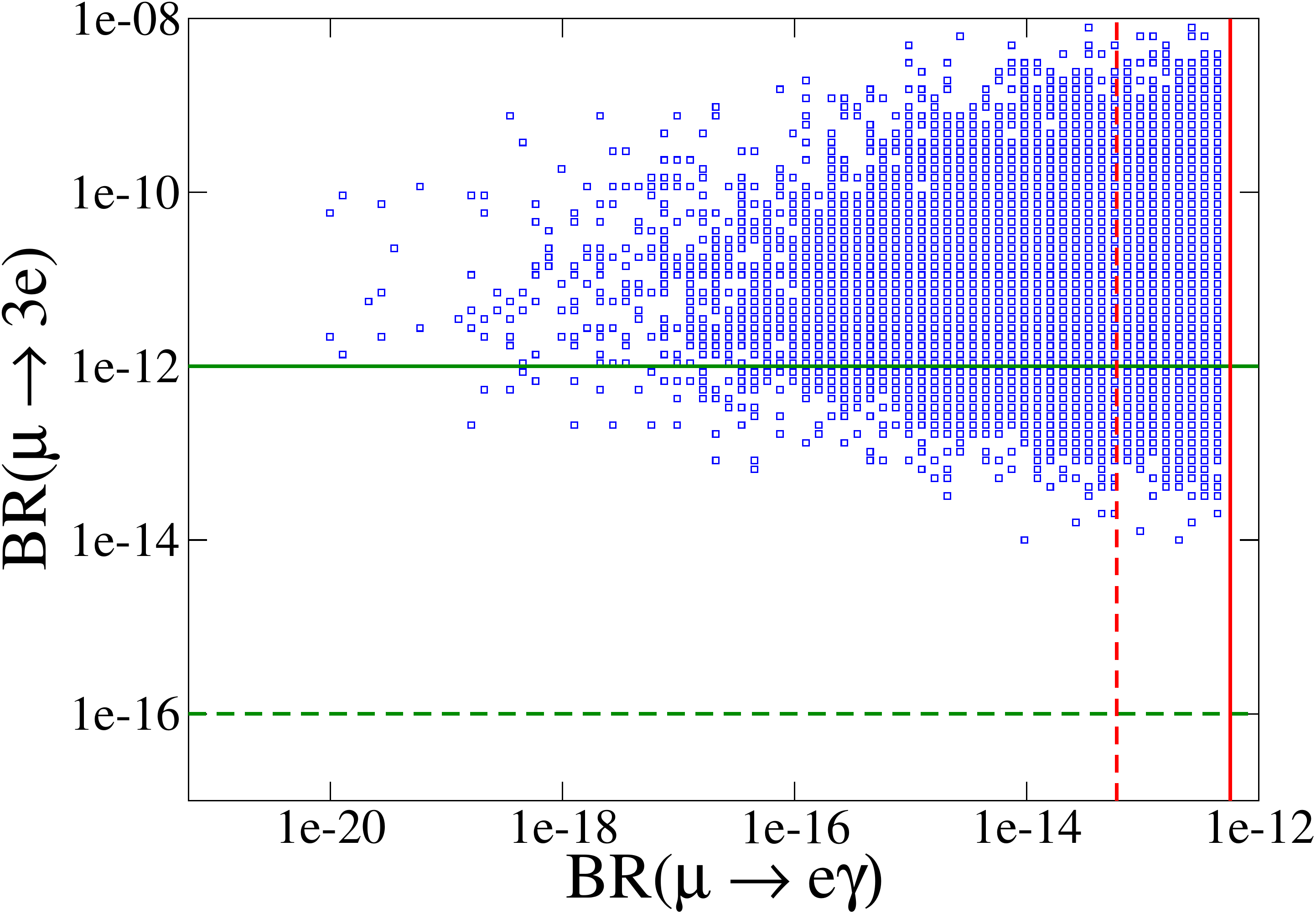} &    \includegraphics[scale=0.23]{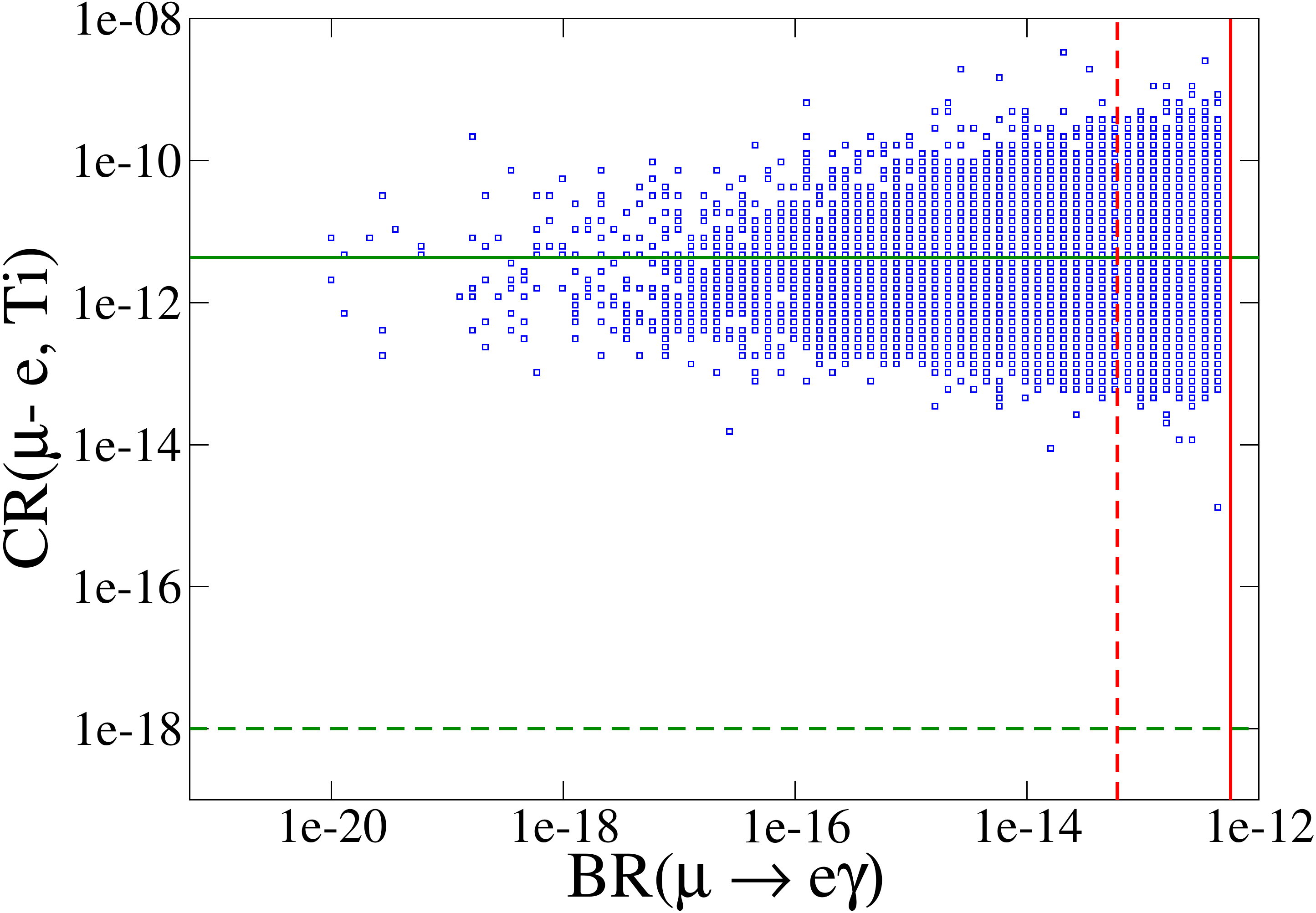} \\
(a) &    (b)
\end{tabular}
\caption{\small (a): Scatter plot of BR($\meg$) versus BR($\mte$) including current bounds (solid lines) and expected future sensitivities (dashed lines). (b): Scatter plot of BR($\meg$) versus CR($\mec$,Ti) including current bounds (solid lines) and future sensitivities (dashed lines).
\label{fig:nocoannmuegvsall} }
\end{figure}

The impact that future LFV searches will have  on the parameter space of this model is illustrated in figure \ref{fig:nocoannmuegvsall}.  It shows scatter plots of BR($\meg$) versus BR($\mte$)  in the left panel and  versus CR($\mec$, Ti) in the right panel.  In addition,  their current bounds (solid lines) and their expected future sensitivities (dashed lines) are also displayed.  The expected improvement on the $\meg$ bound would allow, if no signal is found, to exclude an important region of the parameter space, but many models would remain viable. Not so with $\mte$ and $\mec$ conversion in nuclei.  Future $\mte$ and $\mec$ conversion experiments will probe the entire parameter space consistent with dark matter via $N_1$-$N_1$ annihilations. And they can do so even if they fall short of reaching their expected sensitivity. For $\mte$, a branching ratio sensitivity of $10^{-14}$ would be enough to effectively exclude this scenario, two orders of magnitude larger than what is achievable at the Mu3e experiment. For $\mec$ conversion in Titanium, a  rate sensitivity of $10^{-14}$, four orders of magnitude larger than the most optimistic figure, could  exclude practically the entire parameter space.

Summarizing,   the viable parameter space of the scotogenic model will be probed in different ways by future LFV experiments. Searches for LFV $\tau$ decays can, by themselves, probe a significant part, including the entire region with $M_{N_3}<3\TeV$. Future searches for $\mte$ and $\mec$ conversion in nuclei can go deeper and independently probe the whole parameter space we considered, which extends up to $M_{N_3}= 10\TeV$.  If future LFV experiments fail to find a signal, this scenario, where $N_1$-$N_1$ annihilations set  the dark matter relic density, can be ruled out. 
\subsection{Dark matter via $N_1$-$\eta$ coannihilations}
If the dark matter relic density is obtained via $N_1$-$\eta$ coannihilations, the Yukawa couplings can be smaller and so are the rates for LFV processes. As we will see, future experiments will not be able to exclude this possibility but they can test a significant part of the parameter space. In this section, we first present the viable parameter space and then identify the regions that can and cannot be probed with future LFV experiments.  

The free parameters for the random scan are the same as before, with the condition that the mass splitting between the dark matter particle and the scalars, which in this case we take to be degenerate, must be small. On these parameters, we apply the constraints mentioned in the previous section,  further requiring that the relic density be determined by $N_1$-$\eta$ coannihilations only. We obtained in this way a sample of  $2\times 10^4$ viable models, on which our following analysis is based.

\begin{figure}[t]
\begin{center} 
\includegraphics[scale=0.45]{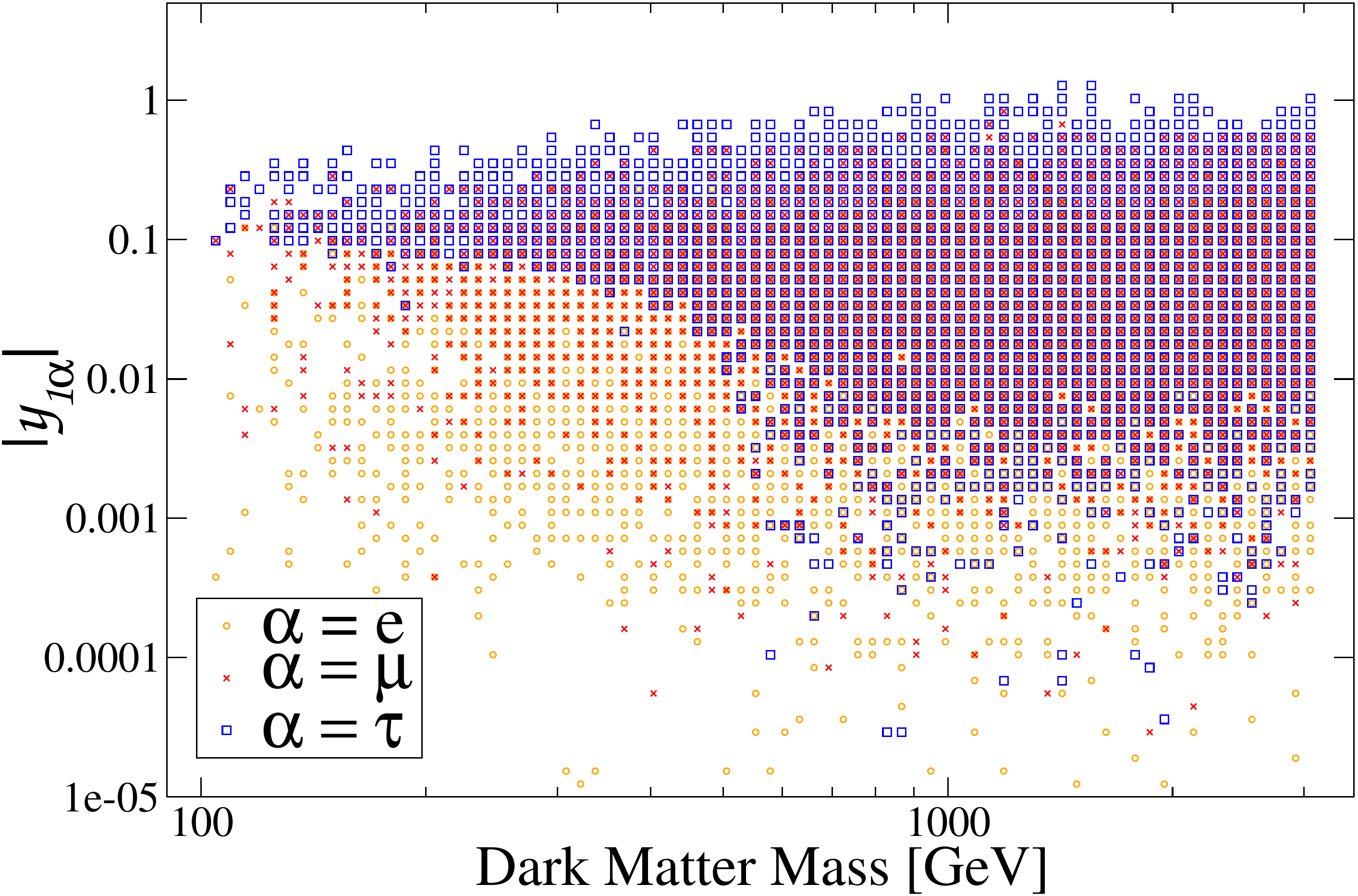}
\caption{\small The Yukawa couplings associated with the dark matter particle, $y_{1\alpha}$, as a function of $\mdm$. In this case, the dark matter relic density is determined by $N_1$-$\eta$ coannihilations.\label{fig:coanndmyuk}}
\end{center}
\end{figure}

To begin with, we project, in figure \ref{fig:coanndmyuk}, the viable models into the plane ($\mdm$, $y_{1\alpha}$) for $i=e,\mu,\tau$. Since  $N_1$-$\eta$ coannihilations  determine the relic density, the mass splitting between $N_1$ and the scalars is necessarily small, and the lower bound on $\mdm$ is now set by the collider bound on the mass of the scalars, $\sim 100\GeV$. At the other end, we see that we can find viable models up to the highest dark matter masses explored in the scan, $\sim 3\TeV$. The dark matter Yukawa couplings are indeed smaller now, never reaching our perturbativity limit. For $\mdm\sim 100\GeV$, the largest value of $|y_{1\alpha}|$ is about $0.3$, and it increases with $\mdm$ until it reaches $1$-$2$ for masses of order $1-3\TeV$. Notice that at low dark matter masses, $\mdm<300\GeV$, $|y_{1\tau}|$ tends to be large, lying on a narrow band between $0.1$ and $0.4$. But at higher masses, it can be much smaller, reaching values below $10^{-3}$ in some cases. In fact, there is no  clear  hierarchy among the different  couplings. One can easily find models where $|y_{1\mu}|>|y_{1\tau}|$ or where $|y_{1e}|>|y_{1\mu}|$. It is true, though, that $|y_{1e}|$ is very rarely the largest one among the three couplings.  Thus, the coannihilation processes  will  feature one lepton from the second or third generation as a final state --e.g. $N_1\eta^+\to W^+\bar \nu_\tau$ or $N_1\eta^0\to W^+\mu^-$.

\begin{figure}[t]
\begin{center} 
\includegraphics[scale=0.45]{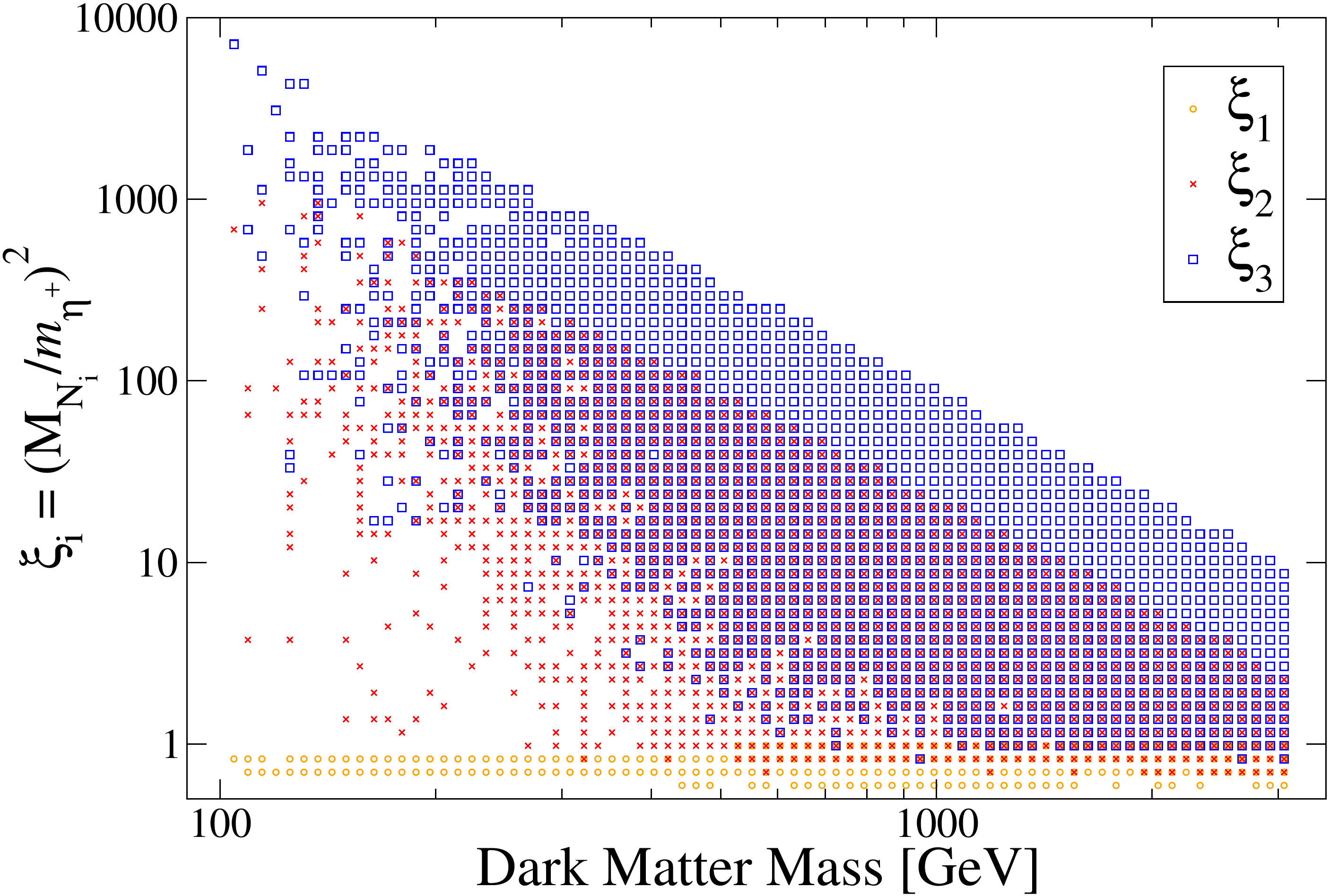}
\caption{\small The parameters $\xi_{i}=(M_{N_i}/m_{\eta^+})^2$ as a function of $\mdm$. In this case, the dark matter relic density is determined by $N_1$-$\eta$ coannihilations.\label{fig:coannxi}}
\end{center}
\end{figure}

The allowed values of $\xi_i$ ($i=1,2,3$) are shown in figure \ref{fig:coannxi}. As a consequence of the coannihilation condition, $\xi_1$ is seen to be constrained to a narrow range just below $1$. $\xi_{2}$ and $\xi_{3}$, on the other hand, tend to be much larger than $1$ and span the whole range explored in the scan.

Regarding LFV $\tau$ decays, we found that in this case they are all too suppressed to probe the model. Their current bounds do not constrain the viable parameter space at all, and only a handful of points feature branching ratios above their expected future sensitivities.  For this reason, we will only examine the $\mec$ transitions in the following. 

\begin{figure}[t]
\begin{center} 
\includegraphics[scale=0.45]{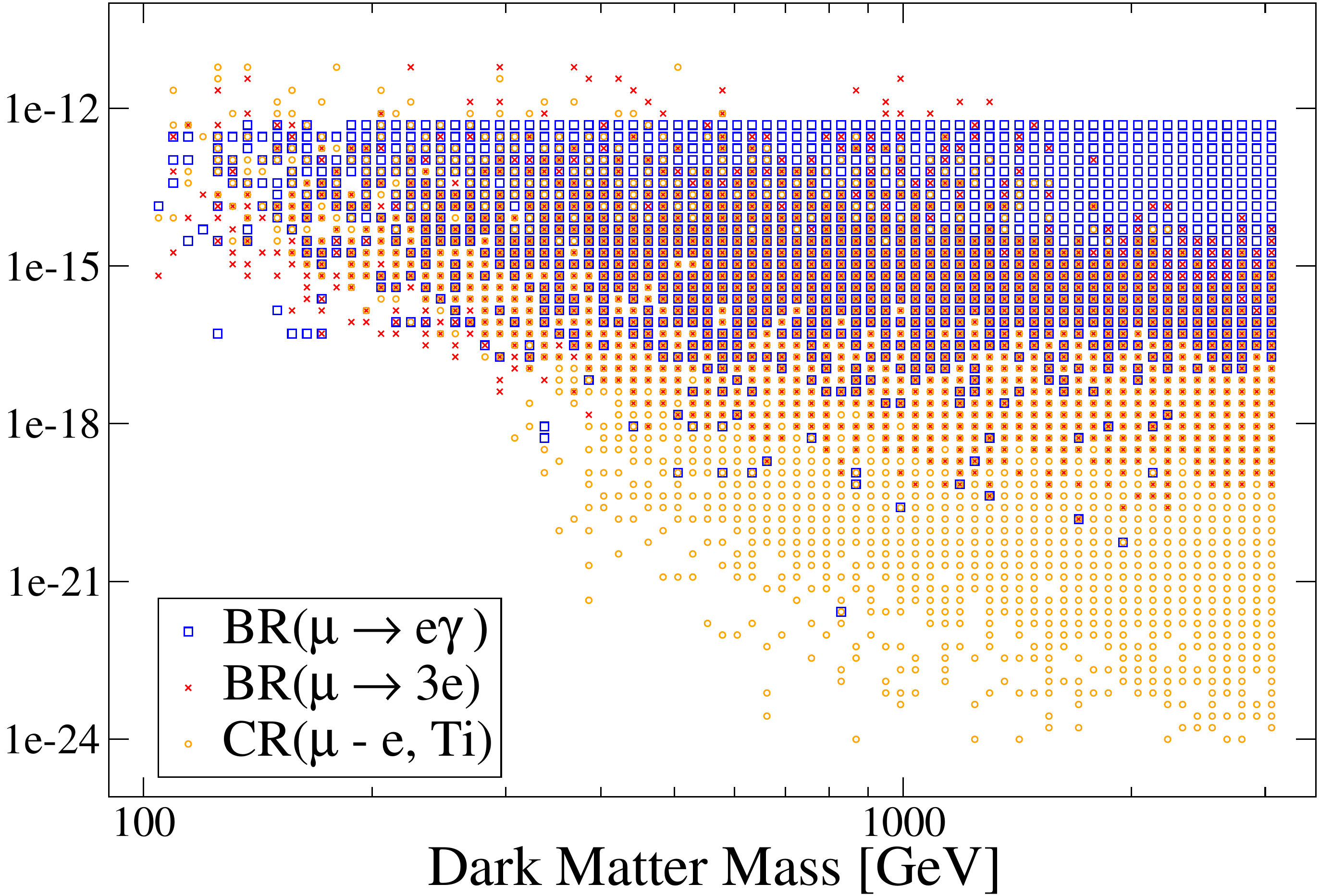}
\caption{\small The rates of $\meg$, $\mte$, and $\mec$ conversion in Titanium as a function of the dark matter mass.  \label{fig:coannalllfv}}
\end{center}
\end{figure}
The predicted rates for the different $\mec$ violating processes are shown in figure \ref{fig:coannalllfv}. By construction, all these points are consistent with the current bound on $\meg$. Even though the bounds on $\mte$ and $\mec$ conversion in nuclei could be violated, we observe that only few points do so. In this case, then, it is true that $\meg$ provides the most stringent bound on LFV processes; once it is satisfied, the other bounds are automatically fulfilled too. Notice that all three processes can be very suppressed. $\meg$ and $\mte$ can both reach branching ratios as low as $10^{-20}$ whereas the $\mec$ conversion rate in Titanium extends down to  values below $10^{-24}$.  It is clear, therefore, that this scenario cannot be entirely probed by future LFV experiments.

\begin{figure}[t]
\begin{tabular}{cc}
 \includegraphics[scale=0.23]{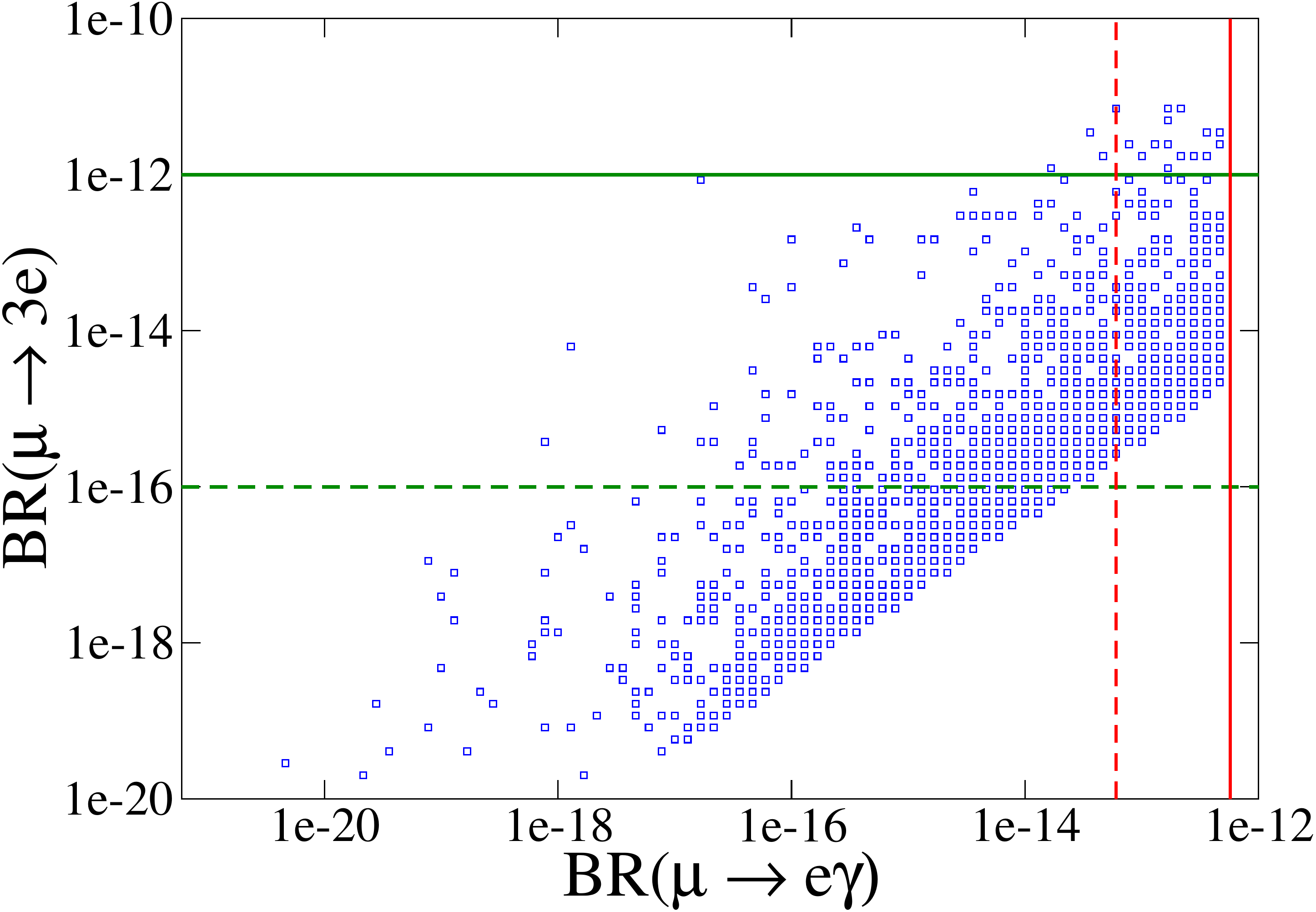} &    \includegraphics[scale=0.23]{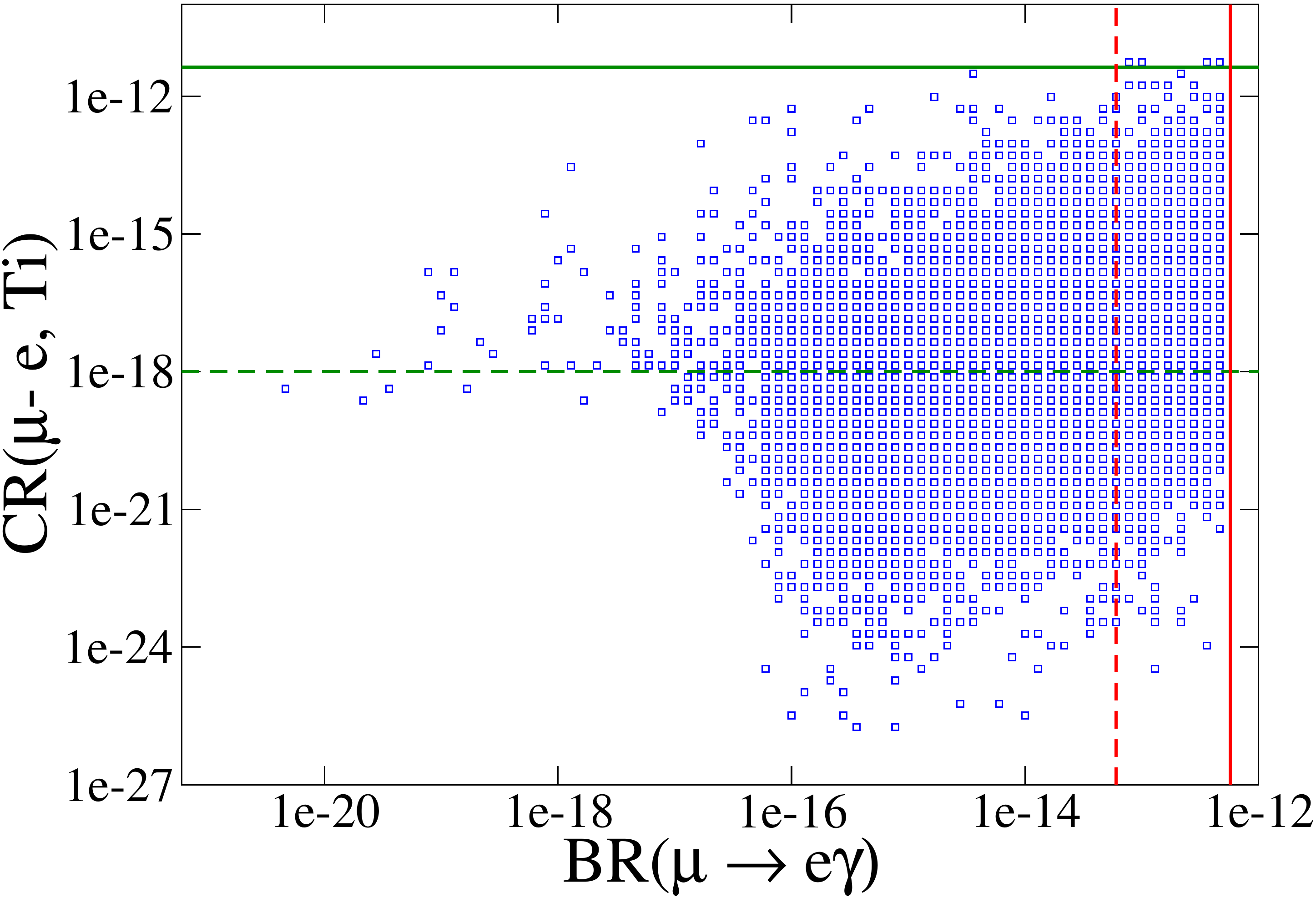} \\
(a) &    (b)
\end{tabular}
\caption{\small (a): Scatter plot of BR($\meg$) versus BR($\mte$) including current bounds (solid lines) and expected future sensitivities (dashed lines). (b): Scatter plot of BR($\meg$) versus CR($\mec$,Ti) including current bounds (solid lines) and future sensitivities (dashed lines). }
\label{fig:coannmuegvsall} 
\end{figure}

This point is further illustrated in figure \ref{fig:coannmuegvsall}, which shows scatter plots of BR($\meg$) versus BR($\mte$), in the left panel, and versus CR($\mec$, Ti) in the right panel. In these figures, the current bounds and the future expected sensitivities are also displayed as solid and dashed lines respectively.  Notice from the left panel that all points lie above a line in this plane determined by the so-called dipole dominance condition ($A_D\gg A_{ND},B$), which implies BR$(\meg)\sim 200$ BR$(\mte)$. Comparing the sensitivity of future experiments, it is clear that the searches for $\mte$ will probe a larger region of the parameter space than the searches for $\meg$. That is, all points with  BR$(\meg)>6\times 10^{-14}$ feature BR$(\mte)>10^{-16}$, but not the other way around. Less conclusive is the result of the comparison between $\meg$ and $\mec$ conversion --see the right panel. In this case there is no clear winner: future $\meg$ experiments will probe regions that cannot be probed with $\mec$ conversion experiments, and viceversa.  In any case,  many viable points will be beyond the expected sensitivity of all future LFV experiments. In the following, we  characterize such points and identify the regions of the parameter space that will be probed by future experiments.

\begin{figure}[t]
\begin{center} 
\includegraphics[scale=0.5]{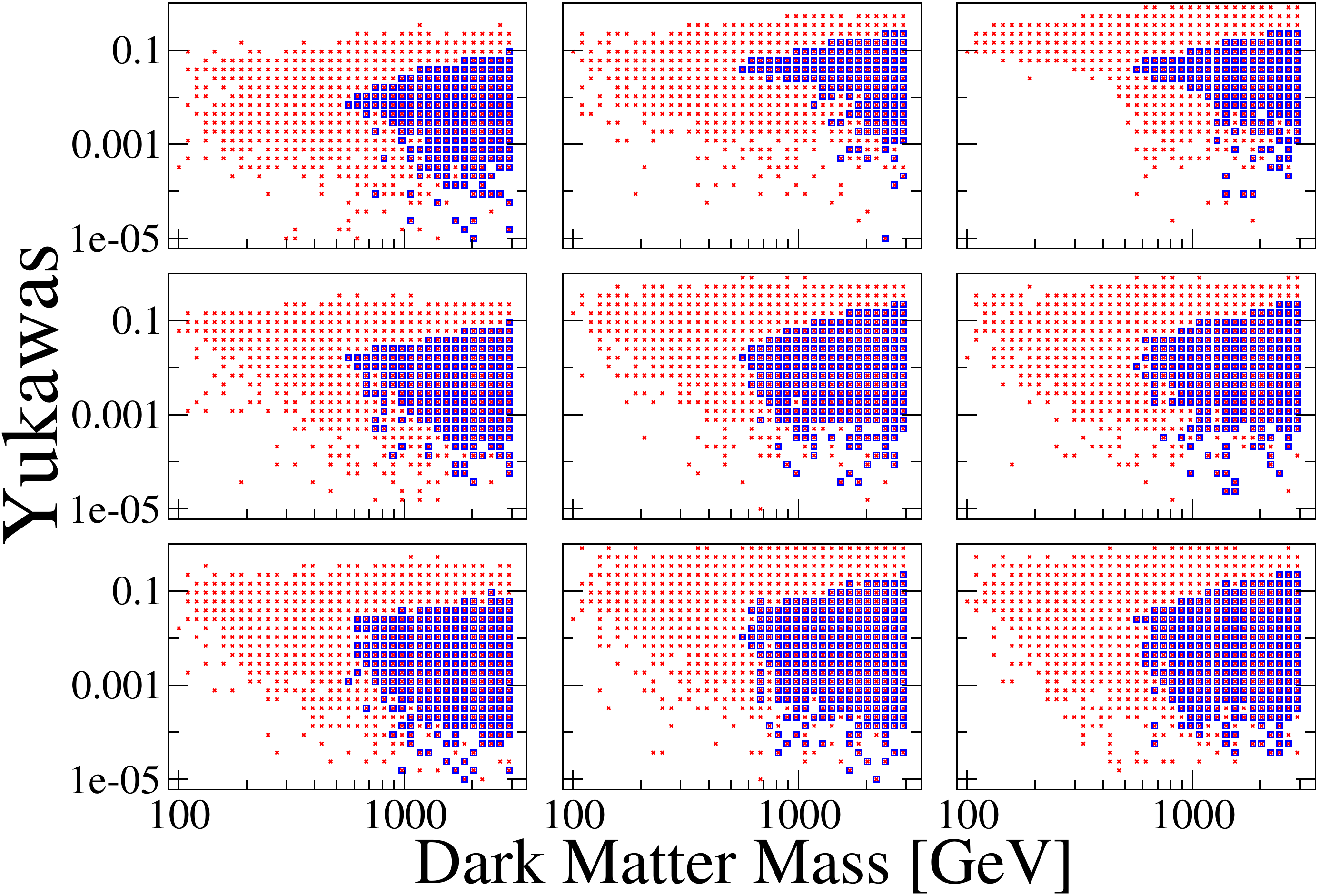}
\caption{\small The values of the Yukawa couplings that are compatible with present bounds (red crosses) and those that are beyond the expected sensitivity of planned LFV experiments (blue squares). Each panel shows the corresponding element of the Yukawa matrix. The top-left panel, for example, shows the element ($1$,$1$) while the bottom-right panel shows the element ($3$,$3$). \label{fig:coannyukarray}}
\end{center}
\end{figure}

Because all lepton flavor violating processes in this model are determined by the neutrino Yukawa couplings, future experiments  have the potential to test the regions featuring  the largest values of $y_{i\alpha}$.  Figure \ref{fig:coannyukarray} illustrates the currently viable region (red crosses) and the regions that lie beyond the expected sensitivity of future LFV experiments (blue squares) for each  element of $|y_{i\alpha}|$. The position of the panel in the array corresponds to the respective matrix element. Thus, the top-left panel shows $y_{1e}$ while the bottom right shows $y_{3\tau}$.  The $x$-axis corresponds instead to the dark matter mass. As expected, the points that lie beyond the expected sensitivity of future LFV experiments feature a heavy spectrum ($\mdm\gtrsim 600\GeV$) and small values of the neutrino Yukawa couplings. The $e$-column of the Yukawa matrix (the first one) is already the most strongly constrained, and future experiments can exclude the region $|y_{i\alpha}|\gtrsim 0.1$ over the entire range of $\mdm$.

\begin{figure}[t]
\begin{center} 
\includegraphics[scale=0.45]{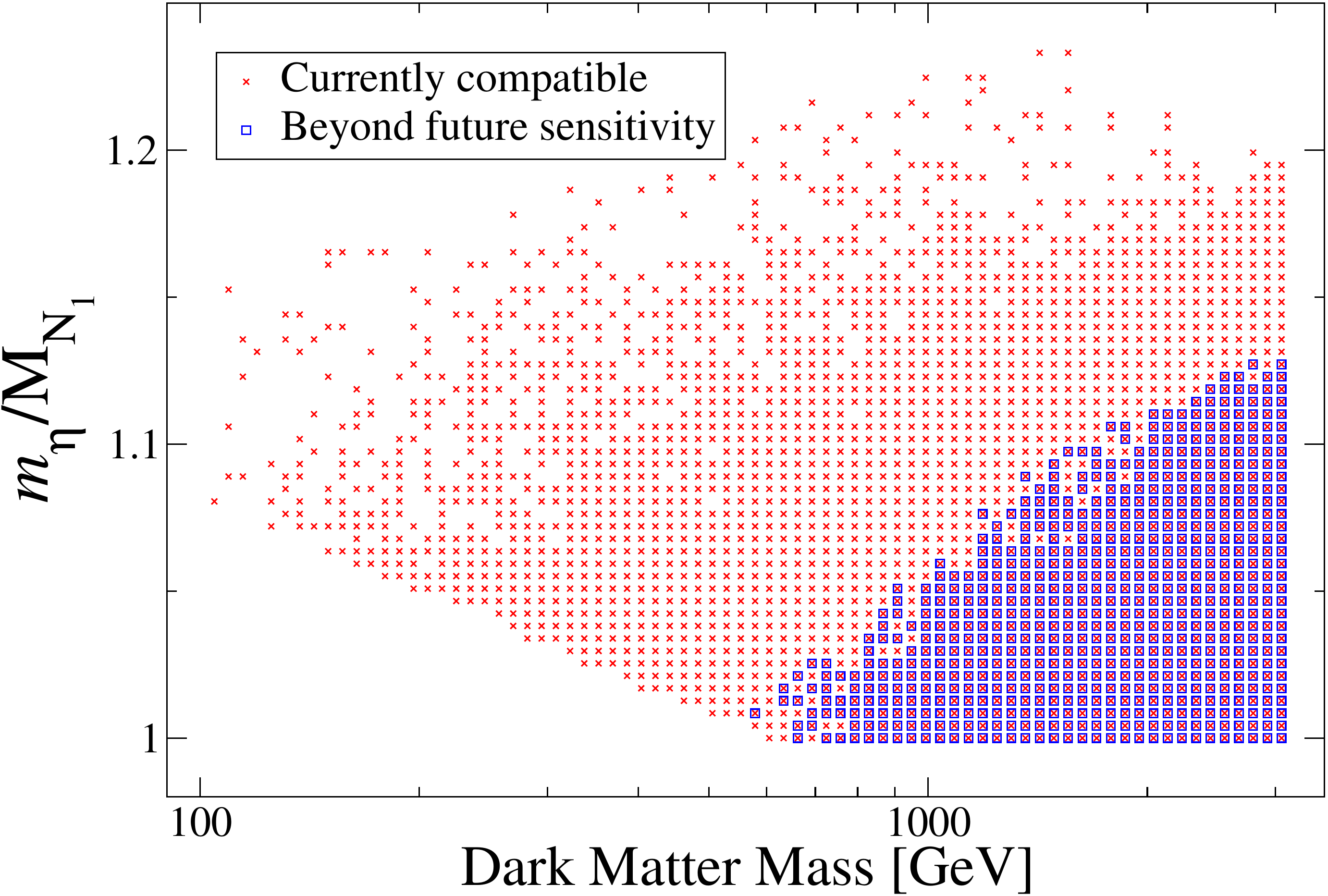}
\caption{\small The regions in the plane ($\mdm$, $\mch/\mdm$) that are compatible with present bounds (red crosses) and those that are beyond the expected sensitivity of planned LFV experiments (blue squares).  \label{fig:coannmassfuture}}
\end{center}
\end{figure}

The mass of the charged scalar ($\mch$) can affect the rates of LFV processes directly and indirectly. On the one hand, $\eta^+$ always appear in the LFV loops, with the result that all form factors are proportional to $1/\mch^2$. If this were the only relevant effect, future experiments could not probe the region where $\mch$ is sufficiently large. But $N_1$-$\eta$ coannihilations imply that $\mch/\mdm$ can never be much larger than one. Moreover, as $\mch$ is increased, the coannihilation effect becomes suppressed and  larger Yukawa couplings are required to satisfy the dark matter constraint. This indirect effect then points in the opposite direction, with future experiments unable to probe the low $\mch$ region.  Figure \ref{fig:coannmassfuture} shows a scatter plot of $\mch/\mdm$ versus $\mdm$,  comparing the viable regions (red crosses) with those that lie beyond the sensitivity of future LFV experiments. As expected, $\mch/\mdm$ is always small, never going above $1.25$. This ratio varies approximately between $1.1$ and $1.15$ at small masses, and between $1$ and $1.25$  at large masses. Future experiments have the potential to exclude the region where $\mch/\mdm$ is large, indicating that the indirect effect mentioned above dominates. The regions $\mch/\mdm\gtrsim 1.10$ and $\mch/\mdm\gtrsim 1.15$, for example, could practically be excluded respectively  for $\mdm\lesssim 1\TeV$ and $\mdm\lesssim 3\TeV$. As observed before, the region of low dark matter mass, $\mdm\lesssim 600\GeV$, can be effectively excluded by future experiments for arbitrary values of $\mch/\mdm$.

\begin{figure}[t]
\begin{center} 
\includegraphics[scale=0.45]{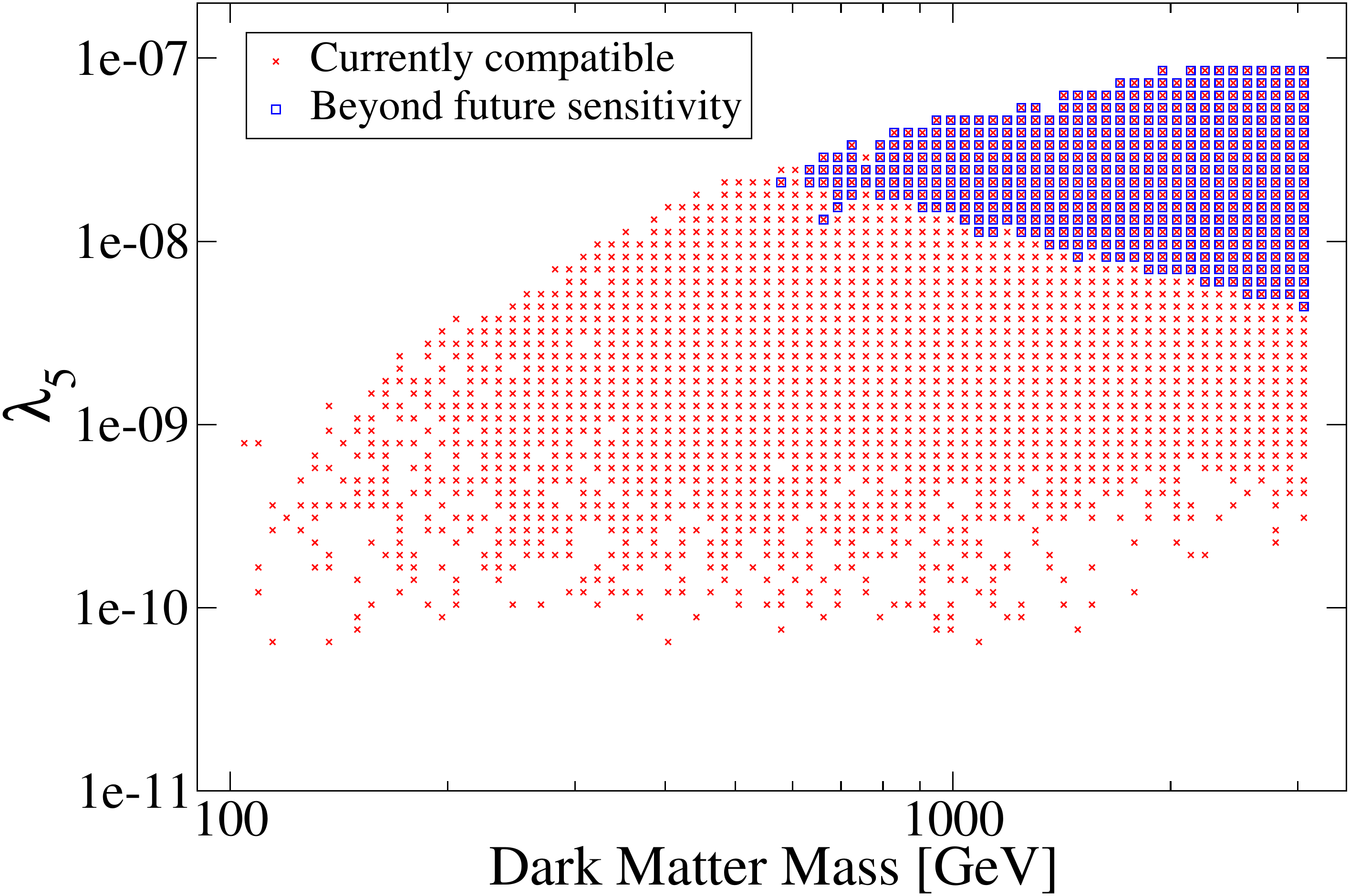}
\caption{\small The regions in the plane ($\mdm$, $\lambda_5$) that are compatible with present bounds (red crosses) and those that are beyond the expected sensitivity of planned LFV experiments (blue squares).  \label{fig:coannfuture}}
\end{center}
\end{figure}

Figure \ref{fig:coannfuture} shows a scatter plot of our models  in the plane ($\mdm,\lambda_5$), again differentiating between all the viable points (red crosses) and those that lie below the expected sensitivity of future experiments (blue squares). $\lambda_5$ is found to vary in this case between $10^{-11}$ and $10^{-7}$. Larger values of $\lambda_5$ would imply, via the neutrino mass constraint, smaller Yukawas, which are not consistent with the requirement of a dark matter density driven by $N_1$-$\eta$ coannihilations. As shown in the figure, future LFV experiments can exclude all models with $\lambda_5\lesssim 10^{-9}$ and practically all models with $\lambda_5<10^{-8}$ and $\mdm\lesssim 600\GeV$. Because the rates for LFV processes do not depend explicitly on $\lambda_5$, this constraint is indirect, via the Yukawa couplings and the neutrino mass scale. In this setup, if future LFV experiments fail to find a signal, $\lambda_5$ should lie within one order of magnitude of $10^{-8}$.

\section{Conclusions}
\label{sec:con}
We demonstrated that future searches for lepton flavor violating processes will have a significant  impact on  the scotogenic model.  Throughout this work, we assumed  the dark matter particle to be the lightest singlet fermion, $N_1$, and considered two cases depending on how its relic density in the early Universe is determined: via self-annihilations or via coannihilations with the odd scalars. For each case we used  a random scan over the entire parameter space of this model to obtain a large sample of viable points, on which our analysis was based.  We found that the prospects to observe a LFV signal strongly depends on the mechanism that sets the  dark matter density   in the early Universe.

When the dark matter density is obtained via $N_1$-$N_1$ annihilations, the Yukawa couplings tend to be large, which in turn translates into significant rates for lepton flavor violating processes. First, we analyzed the resulting parameter space in some detail by projecting the viable points onto different planes. We showed that current bounds on LFV observables strongly constrain the parameter space of the model, although valid regions are still found.  Then we examined the potential of  future LFV experiments.  We demonstrated that searches for $\tau$ decays can probe an important part of the parameter space.  Specifically all models featuring $M_{N_3}\lesssim 3\TeV$ could be excluded if no signal is found. Future searches for $\meg$ will not have a dramatic impact on the parameter space of the model because this decay can be very suppressed. BR$(\mte)$ and CR($\mec$,Ti), on the other hand, typically lie within few orders of magnitude of their present bounds and offer much better prospects. In fact, future searches for $\mte$ and $\mec$ conversion in nuclei will independently
and thoroughly  probe this  region. If no signal of either process were found, the entire parameter space  could be excluded.

When the dark matter density is obtained via $N_1$-$\eta$ coannihilations, the Yukawa couplings need not be so large and consequently the rates for lepton flavor violating processes are smaller. As a result, LFV $\tau$ decays play no role and only $\mec$ processes can test this scenario. We showed that it is possible to  find viable points where  the rates of all these processes are below the expected sensitivity of future experiments. That is, a fraction of the viable models cannot be probed with future experiments. We characterized such models: they feature $\mdm\gtrsim 600\GeV$, small Yukawa couplings, $\mch/\mdm\lesssim 1.15$, and $\lambda_5\gtrsim 10^{-9}$. 

\section*{Acknowledgments}
The authors are grateful to Igor Ivanov for comments on the first version of the manuscript. A.V. is grateful to Takashi Toma for fruitful discussions and acknowledges partial support from the EXPL/FIS-NUC/0460/2013 project financed by the Portuguese FCT. C.Y. is partially supported by the ``Helmholtz Alliance for Astroparticle Physics HAP'' funded by the Initiative and Networking Fund of the Helmholtz Association. 
 
\appendix

\section{Loop functions}
\label{sec:appendix1}
We present in this appendix the loop functions relevant for the computation of the LFV observables. These are
\begin{eqnarray}
F_2(x) &=& \frac{1-6x+3x^2+2x^3-6x^2 \log x}{6(1-x)^4}, \\
G_2(x) &=& \frac{2-9x+18x^2-11x^3+6x^3 \log x}{6(1-x)^4}, \\
D_1(x,y) &=& - \frac{1}{(1-x)(1-y)} - \frac{x^2 \log x}{(1-x)^2(x-y)} -
 \frac{y^2 \log y}{(1-y)^2(y-x)}, \\
D_2(x,y) &=& - \frac{1}{(1-x)(1-y)} - \frac{x \log x}{(1-x)^2(x-y)} -
 \frac{y \log y}{(1-y)^2(y-x)}.
\end{eqnarray}
In the limit $x,y\to1$ and $y\to x$, the functions become 
\begin{eqnarray}
F_2(1)&=&\frac{1}{12},\quad
G_2(1)=\frac{1}{4},\quad
D_1(1,1)=-\frac{1}{3},\quad
D_2(1,1)=\frac{1}{6},
\end{eqnarray}
\begin{eqnarray}
&&D_1(x,x)=\frac{-1+x^2-2x\log{x}}{(1-x)^3},\\
&&D_1(x,1)=D_1(1,x)=\frac{-1+4x-3x^2+2x^2 \log{x}}{2(1-x)^3},\\
&&D_2(x,x)=\frac{-2+2x-(1+x)\log{x}}{(1-x)^3},\\
&&D_2(x,1)=D_2(1,x)=\frac{1-x^2+2x\log{x}}{2(1-x)^3}.
\end{eqnarray}

\bibliographystyle{hunsrt}
\bibliography{darkmatter}

\end{document}